\documentclass[12pt,a4paper]{article}
\usepackage[utf8]{inputenc}

\usepackage{framed}
\usepackage{mdframed}

\usepackage{fullpage}
\usepackage[T1]{fontenc}

\usepackage{epsf}
\usepackage{graphicx}
\usepackage{verbatim}
\usepackage{color}	
\usepackage{bbold}
\usepackage{hyperref}
\usepackage[dvipsnames]{xcolor}
\hypersetup{
     colorlinks=true,
     linkcolor = Blue,
     anchorcolor = Blue,
     citecolor = Blue,
     filecolor = blue,
     urlcolor = Black
     }

\usepackage[most]{tcolorbox}
\usepackage{empheq}
\newtcbox{\mymath}[1][]{%
    nobeforeafter, math upper, tcbox raise base,
    enhanced, colframe=blue!30!black,
    colback=blue!30, boxrule=1pt,
    #1}

\usepackage[american]{babel}
\usepackage{verbatim}
\usepackage[T1]{fontenc}
\usepackage{lmodern}
\usepackage{lipsum}
\usepackage{booktabs}
\usepackage{caption}
\usepackage{cite}
\usepackage{soul,color}
\usepackage[toc,page]{appendix}
\usepackage{float}

\usepackage{longtable}

\usepackage{ifpdf}
\usepackage{cancel}

\usepackage{amsthm, amssymb}
\usepackage{pifont}
\usepackage{bm}
\usepackage{mathtools}
\usepackage{amstext}
\usepackage{braket}
\usepackage{multirow}
\usepackage[normalem]{ulem}
\usepackage{xcolor}
\usepackage{cancel}

\usepackage{slashed}

\newcommand\redsout{\bgroup\markoverwith{\textcolor{red}{\rule[0.5ex]{2pt}{0.4pt}}}\ULon}

\begin{document}
\vspace{5mm}
\vspace{0.5cm}

\def\be{\begin{eqnarray}}
\def\ee{\end{eqnarray}}

\def\ba{\begin{aligned}}
\def\ea{\end{aligned}}

\def\ls{\left[}
\def\rs{\right]}
\def\lc{\left\{}
\def\rc{\right\}}

\def\p{\partial}

\def\S{\Sigma}

\def\s{\sigma}

\def\O{\Omega}

\def\a{\alpha}
\def\b{\beta}
\def\g{\gamma}

\def\ad{{\dot \alpha}}
\def\bd{{\dot \beta}}
\def\gd{{\dot \gamma}}
\newcommand{\ft}[2]{{\textstyle\frac{#1}{#2}}}
\def\ib{{\overline \imath}}
\def\jb{{\overline \jmath}}
\def\Re{\mathop{\rm Re}\nolimits}
\def\Im{\mathop{\rm Im}\nolimits}
\def\trace{\mathop{\rm Tr}\nolimits}
\def\rmi{{ i}}

\newcommand{\dd}{\mathrm{d}}
\newcommand{\vol}{\operatorname{vol}}
\newcommand{\Tr}{\operatorname{Tr}}
\newcommand{\AdS}{\mathrm{AdS}}
\newcommand{\eps}{\varepsilon}
\newcommand{\cF}{\mathcal{F}}
\newcommand{\cV}{\mathcal{V}}
\newcommand{\cO}{\mathcal{O}}
\newcommand{\cT}{\mathcal{T}}
\newcommand{\cQ}{\mathcal{Q}}
\newcommand{\kk}{\kappa_{10}}

\def\N{\mathcal{N}}

\newcommand{\SU}{\mathop{\rm SU}}
\newcommand{\SO}{\mathop{\rm SO}}
\newcommand{\U}{\mathop{\rm {}U}}
\newcommand{\USp}{\mathop{\rm {}USp}}
\newcommand{\OSp}{\mathop{\rm {}OSp}}
\newcommand{\Symp}{\mathop{\rm {}Sp}}
\newcommand{\Sl}{\mathop{\rm {}S}\ell }
\newcommand{\Gl}{\mathop{\rm {}G}\ell }
\newcommand{\Spin}{\mathop{\rm {}Spin}}
\newcommand*\mybluebox[1]{\colorbox{blue!20}{\hspace{1em}#1\hspace{1em}}}

\newlength{\Lnote}
\def\hc{c.c.}

\numberwithin{equation}{section}

\allowdisplaybreaks

\allowbreak



\thispagestyle{empty}
\begin{flushright}

\end{flushright}
\vspace{35pt}

\begin{center}
	    {{\fontsize{16}{20}\selectfont \textbf{Scale-separated AdS$_2$ flux vacua from type II}}}

		\vspace{50pt}

    	{George~Tringas\, and\, Timm Wrase}
\end{center}

		\vspace{30pt}

\begin{center}




Department of Physics, Lehigh University,\\

16 Memorial Drive East, Bethlehem, PA 18015, USA


\vspace{1.3cm}

E-mail: georgios.tringas@lehigh.edu, timm.wrase@lehigh.edu

\vspace{1.3cm}

{ABSTRACT}

\end{center}

Motivated by the question of whether the internal scales of a compactification can be parametrically decoupled from the external curvature scale, and by the possibility of probing such vacua holographically, we construct the first AdS$_2$ flux vacua with parametric scale separation.
We compactify type II string theory on a $G_2$ structure orbifold times a circle in the presence of smeared spacetime-filling O1/O5-planes or O4/O8-planes, together with NSNS and RR fluxes.
We derive the two-dimensional dilaton-gravity effective theory and exhibit families of vacua in which unbounded $F_5$ and $H_7$ fluxes provide parametric control in type IIB, while unbounded $F_4$, $F_8$, and $H_7$ fluxes provide parametric control in type IIA.
For large flux quanta, the string coupling becomes parametrically weak, all internal bulk radii become parametrically large in string units, and the Kaluza--Klein scale separates from the $\mathrm{AdS}_2$ curvature scale.
We analyze the ten-dimensional Killing-spinor equations and identify parametric branches preserving $\mathcal N=(1,1)$ supersymmetry in two dimensions.
We explain how our geometrically scale-separated AdS$_2$ vacua are compatible with a recent no-go theorem for scale separation in theories with extended supersymmetry.
Finally, we discuss various T-dualities leading to other type IIB and type IIA compactifications on $G_2$ structure spaces times a circle and $SU(3)$ structure spaces times a two-torus.

\thispagestyle{empty} 
\setcounter{page}{0}

\baselineskip 6mm 

\newpage


{\hypersetup{hidelinks}
\tableofcontents
}
\pagestyle{empty}

\pagebreak
\pagestyle{plain}
\setcounter{page}{1}

\section{Introduction}\label{sec:introduction}

If string theory describes our universe, the extra spatial dimensions must remain unobserved at low energies. The simplest possibility is that the characteristic length scale of the extra dimensions is much smaller than that of the lower-dimensional effective field theory (EFT), so that the corresponding modes decouple from low-energy physics. In compactifications, denoting these scales by $L_{\rm KK}$ and $L_{\rm EFT}$, respectively, this requires
\begin{equation*}
\frac{L_{\rm EFT}}{L_{\rm KK}}\gg 1 \,.
\end{equation*}
To understand whether such a hierarchy can arise, it is natural to focus on theoretically controlled settings, in particular on solutions protected by supersymmetry, which motivates the study of AdS compactifications and at the same time allows their consistency to be investigated through the AdS$_D$/CFT$_{D-1}$ correspondence.

As part of this program, and motivated by the four-dimensional AdS$_4$ solutions in massive IIA of \cite{DeWolfe:2005uu}, studying whether parametrically controlled, moduli-stabilized, and scale-separated vacua can exist in different dimensions serves two purposes. First, it tests how broadly these properties can be realized in simple string compactifications and helps clarify the underlying mechanisms that make them possible. Second, lower-dimensional AdS vacua may admit simpler holographic duals, providing a more tractable setting in which to investigate their consistency through the AdS/CFT correspondence.
Indeed, in recent years, parametrically controlled and scale-separated AdS$_3$ vacua have been found in several toroidal orbifold compactifications of massive type IIA \cite{Farakos:2020phe,VanHemelryck:2022ynr,Farakos:2025bwf}, type IIB/type I \cite{Arboleya:2024vnp,VanHemelryck:2025qok,Miao:2025rgf,Arboleya:2025lwu,Arboleya:2025jko}, heterotic string theory \cite{Tringas:2025bwe}, and massless type IIA \cite{Tringas:2026ncg}, involving $G_2$ structure spaces or products of $SU(3)$ structure spaces and a circle $S^1$.

Compared with four dimensions, where parametrically controlled classical solutions have so far been realized only in massive type IIA, compactifications to lower dimensions allow these questions to be explored within a broader class of classical string compactifications, offering new perspectives on their consistency.
The first candidate scale-separated vacua preserving extended $\mathcal N=2$ supersymmetry were proposed in $\mathrm{AdS}_3$ in \cite{Cribiori:2026caf}, a result that is especially notable in light of previous arguments constraining scale separation in certain supersymmetric AdS theories with gauged $R$-symmetries \cite{Cribiori:2022trc, Montero:2022ghl}.

The question of whether such properties can arise in two-dimensional compactifications has also attracted attention. 
In particular, \cite{Lust:2020npd} investigated several source-free type IIA $\mathrm{AdS}_2$ product-space solutions and found that, within the classes of internal geometries and flux ansätze considered there, the $\mathrm{AdS}_2$ curvature cannot be parametrically smaller than the curvatures of \emph{all} internal factors. 
A parametrically tunable hierarchy among some curvature scales was subsequently found in type IIB $\mathrm{AdS}_2$ S-fold backgrounds in \cite{Guarino:2024zgq}, although the full internal space does not decouple from the $\mathrm{AdS}_2$ scale and the large-hierarchy limit is accompanied by light Kaluza--Klein modes.
A broader obstruction to scale separation in supersymmetric $\mathrm{AdS}_2$ flux vacua was proposed in \cite{Cribiori:2024jwq}, where the authors argued that flux-supported vacua preserving at least $\mathcal N=(1,1)$ supersymmetry contain fundamental BPS domain-wall particles whose tension is parametrically of the order of the AdS scale, thereby obstructing a parametrically higher ultraviolet cutoff.
More recently, \cite{Cribiori:2026qfs} studied flux compactifications to two dimensions on a toroidal Spin(7) orbifold in massive IIA with fluxes and OF1 sources, finding candidate supersymmetric Minkowski vacua, while flux quantization and tadpole cancellation obstruct full stabilization of the untwisted sector.

The study of the putative CFT duals of scale-separated AdS vacua has also attracted considerable interest, with scale separation corresponding to a parametrically large gap between low-lying operators and Kaluza--Klein modes \cite{Aharony:2008wz, Polchinski:2009ch}.
More recent studies of the putative holographic duals of scale-separated flux vacua uncovered the property of integer conformal dimensions in the scalar spectrum of DGKT-type vacua \cite{Apers:2022zjx, Apers:2022tfm}. These integer dimensions were subsequently related to polynomial shift symmetries in \cite{Apers:2022vfp}. 
More recently, holography has been used to derive quantitative constraints on the spectrum of putative dual CFTs \cite{Perlmutter:2024noo}, as well as consistency conditions on extremal three-point functions \cite{Bobev:2025yxp, Revello:2026eqp, Revello:2026opc, Revello:2026yla}.
The large-flux scaling of several AdS vacua was also reproduced from the near-horizon geometries of D-brane domain walls dual to the unbounded fluxes \cite{Apers:2022vfp}. Building on this brane perspective, more recent work has developed flux-backtracking and explicit brane constructions to reconstruct the configurations underlying scale-separated flux vacua and investigate their decoupling properties \cite{Apers:2025pon, Apers:2026lgi, Arboleya:2026irl}.
Complementary arguments have questioned whether such a decoupling can occur in scale-separated AdS vacua \cite{Bedroya:2025ltj}.

To the best of our knowledge, flux compactifications of type II string theory on seven-dimensional $G_{2}$ structure spaces times a circle with orientifold sources have not previously been studied in two dimensions and we fill this gap here.
The paper is organized as follows.
In Section~\ref{sec:8dinternalspace} we introduce $G_2$ structure spaces and their Joyce orbifold realization. Section~\ref{sec:orientifold} studies the O1/O5 and O4/O8 orientifold systems, their allowed fluxes and tadpole conditions.
Section~\ref{sec:2d-effective-theory} contains the dimensional reduction, the resulting two-dimensional dilaton-gravity theories and the general superpotential.
In Section~\ref{sec:solution} we construct parametrically controlled and scale-separated $\mathrm{AdS}_2$ vacua in type IIB with O1/O5-planes and in massless type IIA with O4/O8-planes, and determine their regimes of classical control.
In Section~\ref{sec:ten-dimensional-supersymmetry} we analyze the ten-dimensional Killing-spinor equations for the type IIB solution, identify an explicit $\mathcal N=(1,1)$ supersymmetric branch, and show that it reproduces the two-dimensional solution.
In Section~\ref{sec:ads2-no-go} we relate the geometric scale separation of the supersymmetric solution to a proposed obstruction for scale-separated $\mathrm{AdS}_2$ vacua with extended supersymmetry.
Section~\ref{sec:Dualities} studies the geometric T-dualities of the type IIB O1/O5 compactification, and Section~\ref{sec:conclusion} concludes.
Appendices~\ref{OPlaneconfigurations}, \ref{2Dequation}, \ref{sec:superpotentialdetails}, and \ref{sec:Tdualitiesdetails} contain the possible orientifold configurations, the two-dimensional vacuum equations, details of the superpotential and scalar-potential derivation, and the explicit T-duality details, respectively.


\section{\texorpdfstring{Eight-dimensional internal space}{}}\label{sec:8dinternalspace}

Compactification of type II string theory to two spacetime dimensions requires an eight-dimensional internal space.
Following the logic of scale-separated compactifications on Calabi--Yau threefolds to four dimensions and on $G_2$ manifolds to three dimensions, a natural first possibility is to consider $\Spin(7)$ manifolds, as in \cite{Cribiori:2026qfs}. 
In the toroidal $\Spin(7)$ orbifold realization considered there, however, the untwisted sector contains a relatively small number of invariant cycles on which fluxes can be supported, substantially restricting the possible flux and source configurations and leading only to Minkowski solutions.

An alternative is to consider eight-dimensional spaces with a product structure, such as $Y_7\times S^1$, with $Y_7$ admitting a $G_2$ structure, or $Y_6\times S^1\times S^1$, with $Y_6$ admitting an $SU(3)$ structure.
The additional globally defined one-forms associated with the $S^1$ factors, together with the invariant forms of the lower-dimensional structure, lead to a richer set of invariant differential forms and hence potentially to a larger flux sector.
In particular, a $G_2$ structure provides the associative three-form $\Phi$ and the coassociative four-form $\Psi$, while an $SU(3)$ structure provides the two-form $J$ and the complex three-form $\Omega$, together with the corresponding higher-degree forms constructed from them.
Product geometries of this type therefore offer greater flexibility for constructing flux-supported two-dimensional vacua, while also providing a richer set of cycles that can support localized orientifold sources, whose interplay with the background fluxes is crucial for moduli stabilization and scale separation in the constructions considered below.


\subsection{\texorpdfstring{$G_{2}$ structure orbifolds times $S^{1}$}{}}

In our explicit construction, we consider an internal geometry of the form
\begin{equation}
X_{8}=Y_{7}\times S^{1},
\end{equation}
where $Y_{7}$ admits a $G_{2}$ structure, while the additional circle provides the globally defined one-form $d\theta$.

Focusing first on the $G_{2}$ sector, the corresponding $G_{2}$ structure forms are the associative three-form $\Phi$ and the coassociative four-form $\Psi=\star_{7}\Phi$, which we take to be
\begin{equation}\label{eq:G2-form}
\begin{aligned}
\Phi
&=
e^{127}-e^{347}-e^{567}
+e^{136}-e^{235}+e^{145}+e^{246},
\\
\Psi
&
=
e^{3456}-e^{1256}-e^{1234}
+e^{2457}-e^{1467}+e^{2367}+e^{1357}.
\end{aligned}
\end{equation}
Here $e^a$, $a=1,\ldots,7$, denote an orthonormal frame of one-forms on $Y_7$, with $e^{abc}\equiv e^a\wedge e^b\wedge e^c$ and analogous notation for higher-degree forms. 

In the present work, we consider seven-dimensional spaces admitting a $G_2$
structure or a further reduction of the structure group to one of its subgroups.
The intrinsic torsion of a general $G_2$ structure is encoded in the torsion
classes $\tau_0$, $\tau_1$, $\tau_2$, and $\tau_3$, defined through
\begin{equation}
d\Phi
=
\tau_0\,\Psi
+
3\tau_1\wedge\Phi
+
\star_7\tau_3\,,
\qquad
d\Psi
=
4\tau_1\wedge\Psi
+
\tau_2\wedge\Phi\,,
\end{equation}
where $\tau_i$ transform in the $\mathbf{1}$, $\mathbf{7}$, $\mathbf{14}$, and $\mathbf{27}$ representations of $G_2$, respectively.

We then take the $G_2$ sector to be the Joyce toroidal orbifold
\cite{Joyce:1996a,Joyce:1996b},
\begin{equation}
Y_{7}
=
\frac{T^{7}}{\Gamma},
\qquad
\Gamma\simeq(\mathbb Z_{2})^{3},
\label{eq:Y7-orbifold}
\end{equation}
and retain only the untwisted sector.
The orbifold group $\Gamma$ is generated by three mutually commuting involutions, denoted by $\Theta_{\alpha}$, $\Theta_{\beta}$, and $\Theta_{\gamma}$, and each generator acts by reflections on four of the seven torus coordinates in such a way that the $G_{2}$ structure forms are preserved
\begin{equation}\label{orbifold}
\begin{split}
\Theta_{\alpha}:\quad
(y^1,\ldots,y^7)
&\longmapsto
(-y^1,-y^2,-y^3,-y^4,y^5,y^6,y^7)\,,\\
\Theta_{\beta}:\quad
(y^1,\ldots,y^7)
&\longmapsto
(-y^1,c_1-y^2,y^3,y^4,-y^5,-y^6,y^7)\,,\\
\Theta_{\gamma}:\quad
(y^1,\ldots,y^7)
&\longmapsto
(c_2-y^1,y^2,c_3-y^3,y^4,-y^5,y^6,-y^7)\,.
\end{split}
\end{equation}
The parameters $c_i\in\{0,\frac12\}$ encode possible half-period shifts along the reflected directions.
The remaining non-trivial elements of $\Gamma$ are
\begin{equation}\label{orbifold2}
\begin{split}
\Theta_{\alpha}\Theta_{\beta}:\quad
(y^1,\ldots,y^7)
&\longmapsto
(y^1,c_1+y^2,-y^3,-y^4,-y^5,-y^6,y^7)\,,\\
\Theta_{\alpha}\Theta_{\gamma}:\quad
(y^1,\ldots,y^7)
&\longmapsto
(c_2+y^1,-y^2,c_3+y^3,-y^4,-y^5,y^6,-y^7)\,,\\
\Theta_{\beta}\Theta_{\gamma}:\quad
(y^1,\ldots,y^7)
&\longmapsto
(c_2+y^1,c_1-y^2,c_3-y^3,y^4,y^5,-y^6,-y^7)\,,\\
\Theta_{\alpha}\Theta_{\beta}\Theta_{\gamma}:\quad
(y^1,\ldots,y^7)
&\longmapsto
(c_2-y^1,c_1+y^2,c_3+y^3,-y^4,y^5,-y^6,-y^7)\,,
\end{split}
\end{equation}
and together with the three generators and the identity, these define the group elements $g\in\Gamma$.
Different choices of the $c_i$ leave the local action on differential forms unchanged but modify the global fixed-point structure. 
For example, the unshifted choice $c_1=c_2=c_3=0$ maximizes the number of fixed loci, while non-vanishing half-shifts can render some orbifold elements freely acting.
Finally, since the orbifold group $\Gamma$ acts non-trivially only on $T^7$, the additional $S^1$ is invariant
\begin{equation}
\theta\longmapsto\theta,
\qquad
d\theta\longmapsto d\theta\,.
\end{equation}

We work in ten-dimensional Einstein frame and make the unwarped metric ansatz
\begin{equation}\label{eq:10d-metric}
ds_{10}^{2}
=
g_{\mu\nu}(x)\,dx^{\mu}dx^{\nu}
+
\sum_{a=1}^{7}r_{a}^{2}(dy^{a})^{2}
+
r_{8}^{2}d\theta^{2}\,,
\end{equation}
where $y^{a}\simeq y^{a}+1$ and $\theta\simeq\theta+1$.
The orthonormal frame on the internal space is
\begin{equation}\label{eq:orthonormal-frame}
e^{a}=r_{a}dy^{a}\,,
\qquad
e^{8}=r_{8}d\theta\,,
\end{equation}
and we normalize the integral one-form on the circle according to
\begin{equation}
\int_{S^{1}}d\theta=1 \,.
\end{equation}
We denote the two-, seven- and eight-dimensional volume forms by
\begin{equation}\label{orientation}
\vol_2=\sqrt{-g_2}\,dx^0\wedge dx^1 \,,
\qquad
\vol_7=\sqrt{g_7}\,dy^{1234567}\,,
\qquad
\vol_8=r_8\,\vol_7\wedge d\theta \,,
\end{equation}
where $g_2\equiv\det g_{\mu\nu}$ and $g_7\equiv\det g_{ab}$, with the orientation fixed by the ordering $y^1,\ldots,y^7, \theta$.

The seven untwisted $G_2$ metric moduli are identified with the volumes of the associative three-cycles appearing in $\Phi$.
We assume that these deformations, as well as the dilaton, depend only on the external coordinates.
Using \eqref{eq:G2-form} and \eqref{eq:orthonormal-frame}, the $G_2$ forms can be expanded in the invariant basis as
\begin{equation}\label{eq:G2-moduli}
\Phi
=
\sum_{i=1}^{7}s^i\Phi_i\,,
\qquad
\Psi
=
\sum_{i=1}^{7}
\frac{\mathcal V_{7}}{s^{i}}\,\Psi^i\,,
\end{equation}
while the corresponding seven- and eight-dimensional internal volumes are
\begin{equation}
\mathcal V_7
=
\int_{T^7}\vol_7
=
\prod_{a=1}^{7}r_a
=
\left(\prod_{i=1}^{7}s^i\right)^{1/3},
\qquad
\mathcal V_8
=
\int_{T^8}\vol_8
=
\mathcal V_7 r_8\,.
\end{equation}
Here $\Phi_i$ form an integral basis of orbifold-invariant three-forms, with dual orbifold-invariant four-forms $\Psi^i$,
\begin{equation}\label{eq:integral-basis}
\begin{aligned}
\Phi_i
={}&
\bigl(
dy^{127},
-dy^{347},
-dy^{567},
dy^{136},
-dy^{235},
dy^{145},
dy^{246}
\bigr),
\\[1mm]
\Psi^i
={}&
\bigl(
dy^{3456},
-dy^{1256},
-dy^{1234},
dy^{2457},
-dy^{1467},
dy^{2367},
dy^{1357}
\bigr),
\end{aligned}
\end{equation}
where, for example, $dy^{127}=dy^1\wedge dy^2\wedge dy^7$. 
We choose the normalization
\begin{equation}\label{eq:basis-normalization}
\int_{T^7}\Phi_i\wedge\Psi^j=\delta_i{}^j\,,
\end{equation}
for which the Hodge star acts as
\begin{equation}\label{eq:7d-hodge}
\star_7\Phi_i
=
\frac{\mathcal V_7}{(s^i)^2}\Psi^i,
\qquad
\star_7\Psi^i
=
\frac{(s^i)^2}{\mathcal V_7}\Phi_i,
\end{equation}
with no sum over $i$.

Restricting to forms invariant under the orbifold group, the untwisted cohomology of $X_{8}=Y_{7}\times S^{1}$ relevant for the reduction is
\begin{equation}\label{eq:cohomology}
\begin{aligned}
H^{1}(X_{8})
&=
\langle d\theta\rangle,
\\
H^{3}(X_{8})
&=
\langle\Phi_i\rangle,
\\
H^{4}(X_{8})
&=
\langle\Psi^i,\,
\Phi_i\wedge d\theta\rangle,
\\
H^{5}(X_{8})
&=
\langle\Psi^i\wedge d\theta\rangle,
\\
H^{7}(X_{8})
&=
\langle\vol_{7}\rangle,
\\
H^{8}(X_{8})
&=
\langle\vol_{7}\wedge \,d\theta\rangle.
\end{aligned}
\end{equation}
There are no untwisted invariant two- or six-forms.
This cohomology determines the possible harmonic flux components in the truncation and will therefore be used below to identify the NSNS and RR fluxes compatible with the orbifold and orientifold projections.


\section{Orientifolds, fluxes, and tadpoles}\label{sec:orientifold}

In this section, we present a general analysis of type II compactifications on $G_2$ structure orbifolds times $S^1$ with O1/O5 and O4/O8 orientifold systems.
We discuss the orientifold involutions and the transformation properties of the invariant basis forms, determine the fluxes compatible with the corresponding orientifold projections, and derive the associated tadpole cancellation conditions.


\subsection{Orientifold involution}\label{subsectionInvolutions}

We consider orientifold projections of the form
\begin{equation}
\mathcal O=\Omega_p\sigma\,,
\end{equation}
where $\Omega_p$ is the worldsheet parity operator and $\sigma$ denotes a geometric involution of the internal space whose fixed loci support the orientifold planes.
We will consider two different choices of $\sigma$, leading respectively to the O1/O5 and O4/O8 orientifold systems.

The number and loci of the resulting O-planes depend on the half-period shifts, which determine which involutions in \eqref{orbifold} and \eqref{orbifold2} possess fixed loci. 
The different choices $c_i\in\{0,\frac12\}$ lead to the orientifold configurations summarized in Table~\ref{tab:Oplane-shifts} of Appendix~\ref{OPlaneconfigurations}.

\subsubsection*{O1/O5 setup}

Our internal space has no 2- or 6-cycles in the bulk that could be wrapped by spacetime-filling O3/O7 orientifold planes, and we therefore first consider the O1/O5 system. 
The spacetime-filling O1-planes extend along the two external directions and are localized at isolated points of the eight-dimensional internal space. 
The corresponding geometric involution acts as a reflection on all eight internal directions,
\begin{equation}
\sigma_{O1}:
(y^1,\ldots,y^7,\theta)
\longmapsto
(-y^1,\ldots,-y^7,-\theta)\,.
\end{equation}
In the orbifold $(T^7/\Gamma) \times S^1$ this automatically leads to O5-planes that fill spacetime and wrap four-dimensional cycles of $T^7/\Gamma$ and are localized at $\theta \in \{0,1/2\}$. Specifically, these arise from composing the $\sigma$ involution with non-trivial elements of the orbifold group, leading to fixed loci arising from the spatial involutions 
\begin{equation}
\sigma_{O5_g} =\sigma g\,, \qquad g\in\Gamma\setminus\{1\}\,.
\end{equation}
Since the elements of $\Gamma$ preserve the $G_2$ invariant spinor, these involutions are compatible with the same supersymmetry as the O1-plane.

Since throughout this work we restrict ourselves to the untwisted sector, the different choices of shifts do not affect the transformation properties of the invariant basis.
The untwisted forms in \eqref{eq:cohomology} that generate our bulk cohomology transform as follows under the O1/O5 orientifold projection
\begin{equation}\label{sigmaforms2}
\sigma^*d\theta=-d\theta\,,
\qquad
\sigma^*\Phi_i=-\Phi_i\,,
\qquad
\sigma^*\Psi^i=+\Psi^i\,,
\end{equation}
and consequently
\begin{equation}\label{sigmaforms3}
\sigma^*(\Phi_i\wedge d\theta)=+\Phi_i\wedge d\theta\,,
\qquad
\sigma^*(\Psi^i\wedge d\theta)=-\Psi^i\wedge d\theta\,,
\qquad
\sigma^*\vol_7=-\vol_7\,.
\end{equation}
The action in equation \eqref{sigmaforms2} preserves half the supersymmetry of the compactification as we will show explicitly below in section~\ref{sec:ten-dimensional-supersymmetry}. So, our orientifold of a $G_2$ structure space times a circle preserves a total of two real supercharges.

\subsubsection*{O4/O8 setup}

In the type IIA compactification, we consider spacetime-filling O8-planes wrapping the entire seven-dimensional space $T^7/\Gamma$ and localized along the additional circle.
The corresponding geometric involution therefore acts as a reflection of the $S^1$ direction,
\begin{equation}
\sigma_{O8}
:
(y^1,\ldots,y^7,\theta)
\longmapsto
(y^1,\ldots,y^7,-\theta)\,,
\end{equation}
with fixed loci at $\theta\in\{0,1/2\}$ supporting the O8-planes.

In the orbifold $(T^7/\Gamma)\times S^1$, composing the O8 involution with the non-trivial elements of the orbifold group gives spacetime-filling O4-planes, which wrap three-dimensional cycles of $T^7/\Gamma$ and are localized in the remaining five internal directions. 
Their geometric involutions are
\begin{equation}
\sigma_{O4_g}
=
\sigma_{O8}g\,,
\qquad
g\in\Gamma\setminus\{1\}\,.
\end{equation}
Since the elements of $\Gamma$ preserve the $G_2$ invariant spinor, these involutions
are compatible with the same supersymmetry as the O8-plane.

Since throughout this work we restrict ourselves to the untwisted sector, the different choices of shifts do not affect the transformation properties of the invariant basis. 
The untwisted forms in \eqref{eq:cohomology} that generate our bulk cohomology transform as follows under the O4/O8 orientifold projection,
\begin{equation}\label{sigmaformsO4}
\sigma^{*}d\theta=-d\theta\,,
\qquad
\sigma^{*}\Phi_i=+\Phi_i\,,
\qquad
\sigma^{*}\Psi^i=+\Psi^i\,,
\end{equation}
and consequently
\begin{equation}\label{sigmaformsO4b}
\sigma^{*}\!\left(\Phi_i\wedge d\theta\right)
=
-\Phi_i\wedge d\theta\,,
\qquad
\sigma^{*}\!\left(\Psi^i\wedge d\theta\right)
=
-\Psi^i\wedge d\theta\,,
\qquad
\sigma^{*}\vol_7
=
+\vol_7\,.
\end{equation}


\subsection{Ten-dimensional bosonic action}

The starting point for our compactifications is the democratic type-II
pseudo-action in ten-dimensional Einstein frame,
\begin{equation}\label{eq:typeII-action}
S
=
\frac{1}{2\kappa_{10}^{2}}
\int d^{10}X\sqrt{-g_{10}}
\left[
R_{10}
-\frac12(\partial\phi)^2
-\frac12 e^{-\phi}|H_3|^2
-\frac14
\sum_p
e^{\frac{5-p}{2}\phi}|F_p|^2
\right]\,,
\end{equation}
where the sum runs over even $p=0,2,4,6,8,10$ in type IIA and over odd $p=1,3,5,7,9$ in type IIB. 
We use the convention
\begin{equation}
|F_p|^2
\equiv
\frac{1}{p!}
F_{M_1\ldots M_p}
F^{M_1\ldots M_p}\,.
\end{equation}
The RR field strengths appearing in \eqref{eq:typeII-action} are not
independent, and the corresponding democratic duality relations are
imposed after varying the pseudo-action. In our conventions, the RR and
NSNS duality relations can be written as
\begin{equation}\label{eq:democratic-duality}
F_{10-p}
=
(-1)^{\frac{(p-1)(p-2)}{2}}e^{\frac{5-p}{2}\phi}\star_{10}F_p\,,
\qquad
H_7
=
e^{-\phi}\star_{10}H_3\,.
\end{equation}

Next, we consider the orientifold sources relevant for our compactification.
We restrict throughout to the closed-string sector and neglect possible open-string degrees of freedom associated with D-branes, setting their worldvolume gauge fields and transverse fluctuations to zero.
In addition, we work in the smeared approximation, replacing the localized source distributions by uniform densities over their transverse internal directions.
The corresponding Einstein-frame DBI and Wess--Zumino actions for a generic spacetime-filling Op-plane family can be written as
\begin{equation}\label{eq:Op-action}
S_{Op}
=
-\sum_i\mu_{Op,i}
\int d^{10}X\sqrt{-g_{10}}\,
e^{\frac{p-3}{4}\phi}\rho_{Op,i}
+
\sum_i\mu_{Op,i}
\int C_{p+1}\wedge j_{9-p,i}\,,
\end{equation}
where $\rho_{Op,i}$ denotes the smeared scalar density and
$j_{9-p,i}$ the corresponding unit-normalized Poincar\'e-dual current.
Since the Op-planes are BPS objects, we have chosen conventions in which their signed tensions equal their RR charges. 
In the democratic formulation, variation with respect to the RR potential $C_{p+1}$ gives the sourced equation of motion for $F_{p+2}$.
Upon imposing the corresponding democratic duality relation in \eqref{eq:democratic-duality}, this equation can equivalently be expressed as the sourced Bianchi identity for the magnetically dual field strength $F_{8-p}$,
\begin{equation}\label{eq:general-Op-Bianchi}
dF_{8-p}
=
H_3\wedge F_{6-p}
+
2\kappa_{10}^{2}
\sum_i\mu_{Op,i}\,j_{9-p,i}\,.
\end{equation}
Here $H_3$ denotes the full ten-dimensional NSNS three-form introduced in \eqref{H3FULL}, and the first term on the right-hand side is absent for $p>6$.

Specializing now to the first case of interest, the O1-planes fill the two-dimensional spacetime and are (before our smearing) point-like in the internal space, while each O5$_i$-plane wraps a four-dimensional internal cycle.
For unit-period internal coordinates, the smeared scalar densities and the associated unit-normalized Poincar\'e-dual currents, with orientations fixed consistently with the source projections, are
\begin{equation}\label{eq:smeared-currents}
\rho_{O1}=\frac{1}{\mathcal V_8}\,,
\qquad
\rho_{O5,i}=\frac{1}{s^i r_8}\,,
\qquad
j_8=dy^{1234567}\wedge d\theta\,,
\qquad
j_{4,i}=-\Phi_i\wedge d\theta\,.
\end{equation}
The O8-planes fill the two-dimensional spacetime and wrap the entire seven-dimensional space $Y_7$, while each O4$_i$-plane wraps a three-dimensional internal cycle
\begin{equation}
\rho_{O8}
=
\frac{1}{r_8}\,,
\qquad
\rho_{O4,i}
=
\frac{s^i}{\mathcal V_7 r_8}\,,
\qquad
j_1=d\theta\,,
\qquad
j_{5,i}
=
-\Psi^i\wedge d\theta\,.
\end{equation}
For Op-planes the total tensions and RR charges are negative, whereas for Dp-branes they are positive,
\begin{equation}\label{tensions}
\mu_{Op}<0\,,
\qquad
\mu_{Dp}>0\,,
\end{equation}
and we denote the net signed source contribution by $\mu_p=\mu_{Op}+\mu_{Dp}$\,, with the corresponding source multiplicities understood to be included in $\mu_{Op}$ and $\mu_{Dp}$.


\subsection{Compatible fluxes and tadpole cancellation}

The fluxes allowed in the presence of orientifold planes are determined by their transformation properties under the corresponding geometric involution. 
The ten-dimensional NSNS and RR field strengths transform according to the standard orientifold parity rules; see \cite{VanRiet:2011yc}. For type II the NSNS three-form is odd under the orientifold involution
\begin{equation}
\sigma^{*}H_3=-H_3\,.
\end{equation}
\subsubsection*{O1/O5 system}
For O1- and O5-planes the relevant RR field strengths transform as
\begin{equation}
\sigma^{*}F_{1}=-F_{1},
\qquad
\sigma^{*}F_{3}=+F_{3},
\qquad
\sigma^{*}F_{5}=-F_{5},
\qquad
\sigma^{*}F_{7}=+F_{7}.
\end{equation}
It follows that the one-form flux can only be expanded along the additional $S^1$, since $d\theta$ is the unique untwisted invariant one-form on the internal space. Therefore,
\begin{equation}\label{F1}
F_1=m\,d\theta\,,
\end{equation}
which has the required odd parity, see \eqref{sigmaforms2}, and is thus compatible with the orientifold projection, with $m$ denoting the corresponding quantized flux.

The ten-dimensional NSNS three-form can contain both an internal and a maximally symmetric electric component,
\begin{equation}\label{H3FULL}
H_3
=
H_3^{\rm int}+H_3^{\rm el}
=
\sum_{i=1}^{7}h_i\Phi_i
+
E\vol_2\wedge\,d\theta\,.
\end{equation}
Both components are odd under $\sigma$, see \eqref{sigmaforms2}, and are therefore compatible with the orientifold projection, with $h_i$ denoting the quantized internal three-form fluxes and $E$ the electric flux parameter associated with the external component.

It will be useful to describe the electric component of $H_3$ in terms of its internal magnetic dual. In ten-dimensional Einstein frame,
\begin{equation}\label{H7H3}
H_7
=
e^{-\phi}\star_{10}H_3^{\rm el}
=
q\,dy^{1234567}\,,
\end{equation}
where $q$ denotes the corresponding quantized flux.
Using the electric part of the NSNS flux in \eqref{H3FULL}, the duality relation gives
\begin{equation}\label{qE}
q
=
e^{-\phi}\frac{\mathcal V_7}{r_8}\,E\,.
\end{equation}
Thus, $E$ and $q$ are not independent but provide equivalent descriptions of the same NSNS flux. In the two-dimensional effective theory we will use the internal fluxes $H_3^{\rm int}$ and $H_7$, while the electric description $H_3^{\rm el}$ will be more convenient for the analysis of the ten-dimensional supersymmetry conditions.

An internal harmonic $F_3$ flux would have to be expanded on the basis forms $\Phi_i$, which are odd under $\sigma$, see \eqref{sigmaforms2}, and since $F_3$ is required to be even, these components are projected out. 
The maximally symmetric component proportional to $\vol_2\wedge\,d\theta$ is also odd, and hence there is no untwisted harmonic $F_3$ background.

For the self-dual five-form, we take the internal magnetic component to be
\begin{equation}\label{F5}
F_5^{\rm mag}
=
\sum_{i=1}^{7}f_i\,\Psi^i\wedge d\theta\,,
\end{equation}
where the $f_i$ denote the corresponding quantized fluxes. Using \eqref{sigmaforms3}, these components have precisely the required odd parity of $F_5$. The complete five-form is obtained by imposing self-duality,
\begin{equation}
F_5
=
F_5^{\rm mag}
+
\star_{10}F_5^{\rm mag}\,.
\end{equation}
Finally, an internal harmonic $F_7$ flux would necessarily be proportional to $dy^{1234567}$, which is odd under the orientifold involution as shown in \eqref{sigmaforms3}. 
Since $F_7$ is even, the harmonic internal $F_7$ flux is projected out.

We now turn to the Bianchi identity in \eqref{eq:general-Op-Bianchi}. 
The electric component of the NSNS three-form does not contribute to the internal tadpole conditions, since $H_3^{\rm el}$, $F_1$, and $F_5^{\rm mag}$ in \eqref{H3FULL}, \eqref{F1}, and \eqref{F5}, respectively, all contain a factor of $d\theta$. Thus, their relevant wedge products vanish identically
\begin{equation}\label{H3F1H3F5}
H_3^{\rm el}\wedge F_1=0\,,
\qquad
H_3^{\rm el}\wedge F_5^{\rm mag}=0\,.
\end{equation}
Moreover, the contribution $H_3^{\rm int}\wedge F_5^{\rm el} \propto \Phi_i\wedge\Phi_j= 0$ and $H_3^{\rm el} \wedge F_5^{\rm el}\propto \vol_2\wedge \vol_2=0$ both vanish.
Consequently, the O5- and O1-plane tadpole conditions involve only $H_3^{\rm int}$ together with the internal RR fluxes $F_1$ and $F_5^{\rm mag}$.

Since the untwisted background fluxes $F_3$ and $F_7$ are absent in the orientifold truncation, the relevant internal Bianchi identities reduce to
\begin{equation}\label{eq:F7-Bianchi-direct}
-H_3^{\rm int}\wedge F_5^{\rm mag}
=
2\kappa_{10}^2\mu_1j_8\,,
\qquad
-H_3^{\rm int}\wedge F_1
=
2\kappa_{10}^2
\sum_{i=1}^{7}\mu_{5,i}j_{4,i}\,.
\end{equation}
Using the flux expansions \eqref{F1}, \eqref{H3FULL}, and \eqref{F5}, integration over the corresponding internal cycles gives
\begin{equation}\label{eq:direct-tadpoles}
\sum_{i=1}^{7}h_i f_i=-2\kappa_{10}^2\mu_1\,,
\qquad
m h_i=+2\kappa_{10}^2\mu_{5,i}\,.
\end{equation}
We note that the electric component of $H_3$, or equivalently its magnetic dual $H_7$, is the only flux considered so far that does not enter the RR tadpole conditions, since the corresponding wedge products vanish identically in \eqref{H3F1H3F5}, and therefore provides a natural unbounded flux direction in the present setup without requiring any tuning among the remaining fluxes.


\subsubsection*{O4/O8 system}

For O4-planes the relevant RR field strengths transform as
\begin{equation}
\sigma^{*}F_{0}=-F_{0}\,,
\quad\,\,
\sigma^{*}F_{2}=+F_{2}\,,
\quad\,\,
\sigma^{*}F_{4}=-F_{4}\,,
\quad\,\,
\sigma^{*}F_{6}=+F_{6}\,,
\quad\,\,
\sigma^{*}F_{8}=-F_{8}\,.
\end{equation}
The Romans mass is projected out, since $F_0$ is required to be odd under the orientifold involution, whereas a constant zero-form is even.
Moreover, there are no untwisted invariant two- or six-forms on $X_8$, and therefore no independent harmonic internal $F_2$ or $F_6$ fluxes.

The four-form sector can in principle be expanded on the two classes of untwisted invariant four-forms as in \eqref{eq:cohomology}.
The first are even under $\sigma$ and are therefore projected out, whereas the second are odd and have precisely the required parity of $F_4$.
Consequently, the surviving internal four-form flux takes the form
\begin{equation}\label{F4O4}
F_4
=
\sum_{i=1}^{7}f_i\,\Phi_i\wedge d\theta\,,
\end{equation}
where the $f_i$ denote the corresponding quantized fluxes. 
The electric six-form components related to \eqref{F4O4} by the ten-dimensional democratic duality relations are fixed by $F_4$ and are not independent.

The eight-form flux can be expanded along the internal volume form,
\begin{equation}\label{F8O4}
F_8
=
Mdy^{1234567}\wedge d\theta\,,
\end{equation}
which is odd under $\sigma$ and is therefore compatible with the orientifold projection. Here $M$ denotes the corresponding quantized flux. 
Its electric dual is likewise fixed by the democratic duality relations and does not constitute an independent flux.

The NSNS sector differs from that of the O1/O5 system. 
The internal three-form flux can only be expanded on the invariant basis as in \eqref{H3FULL}, but since it is even under the O4 involution whereas $H_3$ is required to be odd, the internal harmonic three-form flux is projected out.
The electric component
\begin{equation}\label{H3elO4}
H_3=H_3^{\rm el}
=
E\vol_2\wedge\,d\theta\,,
\end{equation}
is instead odd and survives the projection.
As in the O1/O5 system, this component is expressed in terms of its internal magnetic dual $H_7$ in \eqref{H7H3}, with the corresponding quantized flux $q$ related to the electric flux parameter $E$ through \eqref{qE}.

The O4-planes are magnetic sources for $F_4$. Since there are no untwisted invariant two-forms on which an internal $F_2$ flux could be expanded, the flux-induced term $H_3\wedge F_2$ does not contribute to the internal O4-plane tadpole condition.
Consequently, the O4-plane tadpoles are supported entirely by the non-closure of the four-form flux,
\begin{equation}\label{F4BianchiO4internal}
dF_4
=
2\kappa_{10}^2
\sum_i\mu_{4,i}j_{5,i}\,.
\end{equation}
For the tadpole condition to be satisfied, the corresponding component of $F_4$ must be non-closed and hence non-harmonic. 
For a general $G_2$ structure space, we parameterize the non-closure of the invariant three-forms following \cite{DallAgata:2005zlf} using the symmetric matrix ${\cal M}_{ij}$ defined via 
\begin{equation}
d\Phi_i
=
\sum_j {\cal M}_{ij}\Psi^j\,.
\end{equation}
Using \eqref{F4O4}, the internal Bianchi identity therefore becomes
\begin{equation}\label{O4generaltadpole}
\sum_{i,j}
f_i\,{\cal M}_{ij}
\Psi^j\wedge d\theta
=
2\kappa_{10}^2
\sum_i\mu_{4,i}j_{5,i}\,.
\end{equation}
Only those components of $F_4$ that couple to the non-closed three-forms enter the O4-plane tadpole conditions.
The remaining four-form fluxes, together with $H_7$ and $F_8$, are not constrained by these tadpoles and therefore provide unbounded flux directions.

For the O8 sector, since the Romans mass is projected out and $F_0=0$, the Bianchi identity \eqref{eq:general-Op-Bianchi} requires the net O8/D8 charge to vanish.
We therefore cancel the O8-plane charge and tension locally by introducing coincident D8-branes, so that the O8/D8 sector does not contribute to the remaining tadpole conditions or to the scalar potential.


\section{Two-dimensional effective theory}
\label{sec:2d-effective-theory}

In this section, we derive the two-dimensional effective action obtained by reducing type II supergravity on $X_{8}=Y_{7}\times S^{1}$, including the dilaton and the untwisted geometric moduli in the presence of the invariant fluxes. We work throughout in ten-dimensional Einstein frame.
Since in two dimensions a Weyl rescaling cannot render the coefficient of the Ricci scalar constant, the reduced theory takes the form of a dilaton-gravity theory.


\subsection{Reduction of the Einstein--Hilbert term and potential}

Starting from the Einstein--Hilbert term in the ten-dimensional action \eqref{eq:typeII-action} and using the diagonal metric ansatz \eqref{eq:10d-metric}, the ten-dimensional Ricci scalar takes the form
\begin{equation}
R_{10}
=
R_{2} + R_{7}
-
2\nabla^{2}\log\mathcal V_{8}
-
\sum_{a=1}^{7}(\partial\log r_a)^{2}
-
(\partial\log r_{8})^{2}
-
(\partial\log\mathcal V_{8})^{2} \,.
\end{equation}
Since the determinant of the ten-dimensional metric factorizes as
\begin{equation}
\sqrt{-g_{10}}
=
\sqrt{-g_{2}}\,\mathcal V_{8}\,,
\end{equation}
integration over the internal space and an integration by parts of the
Laplacian term give
\begin{equation}\label{eq:EH-reduced}
S_{\mathrm{EH}}
=
\frac{1}{2\kappa_{10}^{2}}
\int d^{2}x\sqrt{-g_{2}}\,
\mathcal V_{8}
\bigg[
R_{2}+ R_{7}
+
(\partial\log\mathcal V_{8})^{2}
-
\sum_{a=1}^{7}(\partial\log r_{a})^{2}
-
(\partial\log r_{8})^{2}
\bigg]\,,
\end{equation}
and similarly the ten-dimensional dilaton kinetic term reduces to
\begin{equation}
S_{\phi}
=
-\frac{1}{4\kappa_{10}^{2}}
\int d^{2}x\sqrt{-g_{2}}\,
\mathcal V_{8}
(\partial\phi)^{2}.
\label{eq:dilaton-reduced}
\end{equation}

\subsubsection*{O1/O5 setup}
We now turn to the dimensional reduction of the scalar potential. The contributions of the RR fluxes are obtained using the flux expansions in \eqref{F1} and \eqref{F5},
\begin{equation}\label{eq:RRflux-potential}
V_{F_1}
=
\frac12 e^{2\phi}m^2
\frac{\mathcal V_7}{r_8}\,,
\qquad
V_{F_{5,i}}
=
\frac{1}{2r_8\mathcal V_7}f_i^2(s^i)^2 \qquad \text{no sum over }i\,.
\end{equation}
The total five-form contribution is then obtained by summing over the components present in a given model, $V_{F_5}
=
\sum_i V_{F_{5,i}}$.
Each $V_{F_{5,i}}$ includes both the magnetic component and its self-dual electric partner, fixed by the condition $F_5=\star_{10}F_5$.

The NSNS sector contains the internal three-form flux $H_3^{\rm int}$ together with the electric component $H_3^{\rm el}$, whose magnetic dual is $H_7$ that is crucial for obtaining an AdS vacuum. Using the decomposition introduced in \eqref{H3FULL} and the dual description in \eqref{H7H3}, the surviving NSNS fluxes contribute upon dimensional reduction on $Y_7\times S^1$ as
\begin{equation}\label{eq:NSNSflux-potential}
V_{H_{3,i}}
=
\frac12 e^{-\phi}\mathcal V_7 r_8\frac{h_i^2}{(s^i)^2}\,,
\qquad
V_{H_7}
=
\frac12 q^2e^\phi\frac{r_8}{\mathcal V_7}\,.
\end{equation}
Similarly to the $F_5$ sector, the total internal three-form contribution is obtained by summing over the components present in a given model.

Finally, the orientifold sources contribute
\begin{equation}\label{eq:orientifold-potential}
V_{O1/D1}
=
2\kappa_{10}^{2}\mu_{1}
e^{-\phi/2}\,,
\qquad
V_{O5/D5,i}
=
2\kappa_{10}^{2}\mu_{5,i}\,
e^{\phi/2}
\frac{\mathcal V_7}{s^i}\,,
\end{equation}
where the total O5-plane contribution is obtained by summing over the families present in the corresponding orbifold configuration.
The scalar potential arising from the fluxes and sources considered in this compactification takes the schematic form
\begin{equation}
V
=
V_{F_1}
+\sum_i V_{F_{5,i}}
+\sum_i V_{H_{3,i}}
+V_{H_7}
+\sum_i V_{O5/D5,i}
+V_{O1/D1}\,,
\end{equation}
where the sums run over the flux components and source families present in the corresponding model.


\subsubsection*{O4/O8 setup}

We now turn to the dimensional reduction of the scalar potential.
The contributions of the RR fluxes are obtained using the flux expansions in \eqref{F4O4} and \eqref{F8O4},
\begin{equation}\label{eq:RRflux-potential-O4}
V_{F_8}
=
\frac12 e^{-3\phi/2}
\frac{M^2}{r_8\mathcal V_7}\,,\qquad
V_{F_{4,i}}
=
\frac12 e^{\phi/2}
\frac{\mathcal V_7}{r_8}\,
\frac{f_i^2}{(s^i)^2}
\qquad \text{no sum over }i \,.
\end{equation}
The total four-form contribution is obtained by summing over the components present in a given model by $V_{F_4}=\sum_i V_{F_{4,i}}$.
The NSNS sector differs from the type IIB O1/O5 setup because the internal $H_3$ flux is projected out, leaving only its electric component, equivalently described by the magnetic dual $H_7$ as in \eqref{eq:NSNSflux-potential}.

In the type IIA case we also have a non-zero contribution from the integral over the internal Ricci scalar $R_7$ in equation \eqref{eq:EH-reduced}
\begin{equation}
V_{\rm curv} = -\mathcal{V}_8 R_7\,. 
\end{equation}
Finally, the orientifold sources contribute
\begin{equation}\label{eq:orientifold-potential-O4}
V_{O4/D4,i}
=
2\kappa_{10}^{2}\mu_{4,i}\,
e^{\phi/4}s^i\,,
\end{equation}
while the total O4-plane contribution is obtained by summing over the families present in the corresponding orbifold configuration.
The scalar potential arising from the fluxes and sources considered in this compactification takes the schematic form
\begin{equation}
V
= V_{\rm curv} +
\sum_i V_{F_{4,i}}
+
V_{F_8}
+
V_{H_7}
+
\sum_i V_{O4/D4,i}\,,
\end{equation}
where the sums run over the flux components and source families present in
the corresponding model.


\subsection{2D supergravity}

Following the two-dimensional reduction, we formulate the effective theories in the $\mathcal N=(1,1)$ supergravity framework of~\cite{Cribiori:2026qfs}. 
Before orientifolding, a type II compactification preserving a single $G_2$ invariant internal spinor has four real supercharges, corresponding to $\mathcal N=2$ in three dimensions. 
Compactification on the additional circle preserves all four supercharges and therefore gives $\mathcal N=(2,2)$ in two dimensions. 
Since the O1/O5 and O4/O8 planes fill the two-dimensional spacetime and their geometric involutions act only on the internal space, the corresponding orientifold projectors commute with the two-dimensional chirality operator. 
They therefore act separately on the left- and right-moving supercharges. 
In each chirality sector, the orientifold relates the two ten-dimensional supersymmetry parameters and removes one linear combination, reducing $\mathcal N=(2,2)$ to $\mathcal N=(1,1)$. 
For the O1/O5 system this counting is verified explicitly in Section~\ref{sec:ten-dimensional-supersymmetry}.

The dimensionally reduced action takes the form
\begin{equation}
S_{2}
=
\frac{1}{2\kappa_{10}^{2}}
\int d^{2}x\,\sqrt{-g_{2}}
\left[
\frac12 J(\varphi)R_{2}
+
2K_{IJ}(\varphi)\,
\partial_{\mu}\varphi^{I}\partial^{\mu}\varphi^{J}
-
V(\varphi)
\right],
\label{eq:2d-sugra-action}
\end{equation}
where $\varphi^{I}$ collectively denote the scalar fields. For the purpose of identifying the scalar field-space metric, it is convenient to introduce the logarithmic coordinates
\begin{equation}\label{moduli1}
x^i\equiv\log s^i,
\qquad
u\equiv\log r_8,
\end{equation}
and parameterize the scalar fields as
\begin{equation}\label{moduli2}
\varphi^I
=
\left(
x^1,\ldots,x^7,u,\phi
\right).
\end{equation}
We nevertheless keep $s^i$ and $r_8$ as the geometric moduli throughout the rest of the analysis.

The function $J(\varphi)$ determines the coupling to the two-dimensional Ricci scalar, $K_{IJ}(\varphi)$ is the scalar field-space metric, and $V(\varphi)$ is the scalar potential. Comparing with the dimensionally reduced action, we obtain
\begin{equation}\label{J}
J(\varphi)=2\mathcal V_8\,,
\end{equation}
while the non-vanishing components of the scalar field-space metric are
\begin{equation}\label{KIJ}
K_{ij}
=
\frac{\mathcal V_8}{12}
\left(1-3\delta_{ij}\right),
\qquad
K_{iu}=K_{ui}
=
\frac{\mathcal V_8}{6},
\qquad
K_{\phi\phi}
=
-\frac{\mathcal V_8}{4},
\end{equation}
with $K_{uu}=K_{i\phi}=K_{u\phi}=0$.
The inverse field-space metric is defined by $K_{IL}K^{LJ}=\delta_I{}^J$ and is obtained by inverting the full scalar kinetic matrix.

The scalar potential is determined by the field-space metric and the real superpotential
\begin{equation}\label{eq:2d-superpotential-potential}
V
=
-\frac18 K^{IJ}W_IW_J
+
\frac18
\frac{
\left(
4W-J_IK^{IJ}W_J
\right)^2
}{
J_IK^{IJ}J_J
}\,,
\end{equation}
where
\begin{equation}
W_I\equiv\frac{\partial W}{\partial\varphi^I},
\qquad
J_I\equiv\frac{\partial J}{\partial\varphi^I}\,.
\end{equation}

To derive the superpotentials relevant for our compactifications, we first use the scalar potential to determine the required dilaton and volume dependence of the individual flux contributions and then organize them according to the linear structure of the superpotential. For the examples studied below, this procedure leads to a general real superpotential that reproduces the corresponding scalar potentials and can be written in the unified form
\begin{align}\label{WIIgeneral}
W_{\rm II}
=
\int_8
\Bigg[
&
\left(
\frac12\Phi\wedge d\Phi
+
e^{-\phi/2}H_3\wedge\Psi
+
e^{\phi/2}H_7
\right)\wedge e^8
+
\sum_p
e^{\frac{5-p}{4}\phi}\,
\widetilde{\mathcal C}_{8-p}\wedge F_p
\Bigg] \,,
\end{align}
where the sum runs over odd $p$ in type IIB and even $p$ in type IIA, with $\widetilde{\mathcal C}_{8-p}$ denoting the corresponding components of the odd and even calibration polyforms,
\begin{equation}\label{polyforms}
\begin{split}
\widetilde{\mathcal C}_{\rm odd}^{G_2}
&=
\widetilde{\mathcal C}_1+\widetilde{\mathcal C}_3+\widetilde{\mathcal C}_5+\widetilde{\mathcal C}_7
=
e^8+\Phi-e^8\wedge\Psi-\vol_7\,,
\\[2mm]
\widetilde{\mathcal C}_{\rm even}^{G_2}
&=
\widetilde{\mathcal C}_0+\widetilde{\mathcal C}_4+\widetilde{\mathcal C}_8
=
1+\left(\Psi-e^8\wedge\Phi\right)+\vol_8\,.
\end{split}
\end{equation}
The terms in parentheses in \eqref{WIIgeneral} encode the geometric and NSNS contributions and take the same form in both theories, while the final term contains the corresponding RR flux contributions. For a torsion-free $G_2$ structure space, $d\Phi=d\Psi=0$, and the geometric contribution in \eqref{WIIgeneral} vanishes. In the presence of metric flux, the geometry generally carries intrinsic torsion, with the exterior derivatives of the defining forms determined by the corresponding torsion classes.


\section{Vacuum solutions}\label{sec:solution}

In this section, we first derive the general supersymmetry conditions for the two orientifold systems with O1/O5-planes and O4/O8-planes\footnote{Although the O1/O5 and O4/O8 orientifold systems are related at the level of their geometric involutions by T-duality, the two explicit solutions considered here are not T-dual to each other. In particular, the T-dualities required to map the O1/O5 source configuration to the O4/O8 one act along multiple legs of the non-vanishing $H_3$ flux and therefore generate non-geometric fluxes while our two explicit setups are geometric. The geometric T-dualities of the type IIB solution are analyzed in Section~\ref{sec:Dualities} and Appendix~\ref{sec:Tdualitiesdetails}.}. 
We then construct analytic examples whose main distinction, and the motivation for considering both, lies in the different mechanisms of tadpole cancellation.
The tadpole conditions can be satisfied in multiple ways as discussed in \cite{Tringas:2025uyg}. 
Interestingly, previous studies \cite{Farakos:2025bwf} suggest that the way in which tadpoles are canceled in the gravitational theory can affect the conformal dimensions of the holographic dual.
In several scale-separated $\mathrm{AdS}_4$ \cite{Apers:2022tfm} and $\mathrm{AdS}_3$ constructions \cite{Arboleya:2024vnp,VanHemelryck:2025qok,Tringas:2025bwe}, conformal dimensions have been found to be integers when parametric control is achieved through fluxes that remain unbounded without requiring cancellations among large flux contributions to the tadpole conditions. 
By contrast, constructions in which unbounded flux directions require a tuning among flux contributions to the tadpole conditions can lead to non-integer conformal dimensions, as for example in \cite{Farakos:2020phe,Miao:2025rgf}.
Motivated by this distinction, we construct two examples realizing these two different mechanisms of tadpole cancellation, for which the conformal dimensions of the dual theory are indeed non-integer and integer respectively.
This provides a useful setting for future work to investigate whether the appearance of integer or non-integer conformal dimensions is tied to the mechanism of scale separation, and whether recent holographic constraints such as those of \cite{Bobev:2025yxp} can be satisfied universally across dimensions.


\subsection{The O1/O5 system}\label{ssec:O1O5}

The real superpotential reproducing the scalar potential of the compactification is
\begin{equation}\label{eq:superpotential-forms}
W
=
\int_{8}
\left[
e^{-\phi/2}
\left(
H_3^{\rm int}\wedge\Psi+e^\phi H_7
\right)\wedge e^8
+
\Phi\wedge F_5^{\rm mag}
-
e^\phi\vol_7\wedge\,F_1
\right]\,.
\end{equation}
Substituting \eqref{eq:superpotential-forms} into \eqref{eq:2d-superpotential-potential} reproduces the flux contributions obtained from the dimensional reduction 
\begin{equation}\label{eq:potential-forms}
\begin{split}
V={}
\frac12\int_{8}
&\Big[
e^{-\phi}H_3\wedge\star_8 H_3
+
e^{\phi}H_7\wedge\star_8 H_7
+
e^{2\phi}F_1\wedge\star_8 F_1
+
F_5\wedge\star_8 F_5
\Big]
\\[1mm]
&
+e^{\phi/2}
\int_{8}
\Psi\wedge H_3\wedge F_1
-
e^{-\phi/2}
\int_{8}
H_3\wedge F_5\,.
\end{split}
\end{equation}
The cross terms involving the internal $H_3$ flux and the RR fluxes $F_1$ and $F_5$ reproduce, upon using the corresponding Bianchi identities in \eqref{eq:F7-Bianchi-direct}, the contributions of the smeared O5- and O1-plane sources, respectively, as shown in Appendix~\ref{sec:superpotentialdetails}.

Before turning to specific flux choices, we can integrate the superpotential in \eqref{eq:superpotential-forms} explicitly over the internal space and obtain the general superpotential of the toroidal compactification in terms of the fluxes and geometric moduli,
\begin{equation}\label{eq:superpotential}
W
=
e^{-\phi/2}\mathcal V_8
\sum_{i=1}^{7}\frac{h_i}{s^i}
+
e^{\phi/2}qr_8
+
\sum_{i=1}^{7}f_i s^i
-
e^\phi m\mathcal V_7 \,.
\end{equation}
The supersymmetry conditions are \cite{Cribiori:2026qfs}
\begin{equation}\label{W0}
W=0\,,
\qquad
W_I= \frac{
J_LK^{LM}W_M-4W
}{
J_LK^{LM}J_M
} J_I\,.
\end{equation}
Since
\begin{equation}
J_i=\frac{2}{3}\mathcal V_8,
\qquad
J_u=2\mathcal V_8,
\qquad
J_\phi=0\,,
\end{equation}
they are equivalently expressed as
\begin{equation}\label{W00}
W=0,
\qquad
W_\phi=0,
\qquad
W_i=\frac13W_u,
\qquad
i=1,\ldots,7\,.
\end{equation}
The condition $W_\phi=0$ first relates the total $H_3$ contribution to those of $F_1$ and $H_7$, while $W=0$ then determines the total $F_5$ contribution.
The seven conditions $W_i=\frac13W_u$ determine the corresponding relations between the individual $H_3$ and $F_5$ components. 
Summing these equations and using the previous two conditions fixes the $H_7$ contribution, giving the following system
\begin{equation}\label{eq:BPS-general}
\begin{split}
e^{-\phi/2}\mathcal V_8\frac{h_i}{s^i}
-f_i s^i
&=
-\frac{1}{2}e^\phi m\mathcal V_7 \,,
\qquad \text{no sum over }i\,,\\
\sum_{i=1}^{7}f_i s^i
&=
2e^\phi m\mathcal V_7 \,,
\\
e^{\phi/2}q r_8
&=
\frac{1}{2}e^\phi m\mathcal V_7\,.
\end{split}
\end{equation}


\subsubsection{Example I: Classical solution with scale separation}

In order to solve this system of equations, we must first select a specific orientifold configuration and specify the corresponding flux ansatz.
As discussed in subsection \ref{subsectionInvolutions}, the different configurations have the same action on the untwisted invariant basis, which is the sector retained throughout our analysis, although their fixed-locus and source structures differ. 
For definiteness, we focus on the configuration  $\mathcal C_{4b}$\footnote{At the level of the untwisted two-dimensional vacuum equations, $\mathcal C_{4a}$ and $\mathcal C_{4b}$ lead to the same class of solutions, up to a relabeling of the $G_2$ moduli.
They nevertheless correspond to different ten-dimensional orbifold and orientifold backgrounds.
This distinction becomes particularly relevant in the T-duality analysis.
Although the vacuum physics of the two configurations is equivalent within the untwisted sector, their networks of geometric T-dualities are different. 
This is because T-duality is sensitive to the specific coordinate triples supporting the non-vanishing components of $H_3^{\rm int}$, rather than only to the number of non-zero flux coefficients $h_i$.}.
As we will show below, in this case the moduli can be stabilized without requiring a non-vanishing net D-brane contribution to the scalar potential, with the source sector contributing only through a net orientifold term.

For the configuration $\mathcal C_{4b}$ in Table \ref{tab:Oplane-shifts}, the
geometric involutions associated with the four $O5$-plane families act on the
internal coordinates of the eight-dimensional space as
\begin{equation}
\begin{split}
\sigma_{\mathrm{O5}_{\alpha}}
=\sigma_{\mathrm{O1}}\Theta_{\alpha}:\quad
(y^{1},\ldots,y^{7},\theta)
&\longmapsto
(y^{1},y^{2},y^{3},y^{4},-y^{5},-y^{6},-y^{7},-\theta)\,,
\\[1mm]
\sigma_{\mathrm{O5}_{\beta}}
=\sigma_{\mathrm{O1}}\Theta_{\beta}:\quad
(y^{1},\ldots,y^{7},\theta)
&\longmapsto
(y^{1},y^{2},-y^{3},-y^{4},y^{5},y^{6},-y^{7},-\theta)\,,
\\[1mm]
\sigma_{\mathrm{O5}_{\alpha\beta}}
=\sigma_{\mathrm{O1}}\Theta_{\alpha}\Theta_{\beta}:\quad
(y^{1},\ldots,y^{7},\theta)
&\longmapsto
(-y^{1},-y^{2},y^{3},y^{4},y^{5},y^{6},-y^{7},-\theta)\,,
\\[1mm]
\sigma_{\mathrm{O5}_{\alpha\gamma}}
=\sigma_{\mathrm{O1}}\Theta_{\alpha}\Theta_{\gamma}:\quad
(y^{1},\ldots,y^{7},\theta)
&\longmapsto
\left(
\tfrac12-y^{1},y^{2},\tfrac12-y^{3},y^{4},
y^{5},-y^{6},y^{7},-\theta
\right)\,.
\end{split}
\end{equation}
while the corresponding smeared transverse currents are
\begin{equation}
j_{4,\alpha}
=-\Phi_3\wedge d\theta \,,\quad\quad
j_{4,\beta}
=-\Phi_2\wedge d\theta \,,\quad\quad
j_{4,\alpha\beta}
=-\Phi_1\wedge d\theta \,,\quad\quad
j_{4,\alpha\gamma}
=-\Phi_4\wedge d\theta \,.
\end{equation}
To cancel the four O5-plane tadpoles associated with the relevant cycles, we take
\begin{equation}\label{eq:H3-ansatz}
(h_1,\ldots,h_7)
=
(-h,-h,-h,-h,0,0,0)\,.
\end{equation}
The tadpole conditions in \eqref{eq:direct-tadpoles} imply \(m h_i<0\) for \(i=1,\ldots,4\), and we therefore take \(m>0\) and write \(h_i=-h\) with \(h>0\).
Since $r_8$ and $\mathcal V_7$ are positive, the last equation in \eqref{eq:BPS-general} fixes $q$ to have the same sign as $m$. 
Thus, for the choice $m>0$ made here, the supersymmetric solution has $q>0$.
The O5-plane charges can be canceled by an appropriate choice of fluxes, as in the analogous spaces studied in \cite{Miao:2025rgf}, although we keep $h_i$ and $m$ general in the following analysis.

Therefore, the fluxes $F_1$ and $H_3^{\rm int}$ are bounded by the O5-plane tadpole conditions. 
By contrast, as discussed in the previous section, the electric component $H_3^{\rm el}$ does not enter the RR tadpoles. 
Since $H_3^{\rm el}$ contains a leg along $d\theta$, its wedge product with both $F_1$ and $F_5^{\rm mag}$ vanishes identically. 
It is therefore unconstrained by tadpole cancellation.

As we will see below, obtaining the desired parametrically controlled regime requires at least one unbounded $F_5$ flux direction. 
For simplicity, we cancel the total O1-plane charge and tension by introducing space-filling D1-branes so that the remaining O1/D1 tadpole condition requires the flux contribution to vanish, i.e. we have $\mu_1=0$ in \eqref{eq:F7-Bianchi-direct} and \eqref{eq:direct-tadpoles}.
A convenient flux choice is\footnote{The tadpole condition allows the more general anisotropic choice $(f_1,\ldots,f_7)=(-N_2-N_3-N_4,N_2,N_3,N_4,G,M,R)$.
Allowing these flux parameters to scale independently introduces additional anisotropy in the internal geometry, since the different fluxes thread different cycles and can therefore stabilize the corresponding moduli at different values.
Since this additional freedom is not essential for the class of solutions studied here, we restrict to $(N_2,N_3,N_4)=(N,-N,-N)$, and
$G=M=R=N$, which reproduces the flux choice used below.
We note, however, that the more general anisotropic ansatz may allow additional parametric regimes with different scalings of the internal radii and the string coupling.}
\begin{equation}\label{eq:F5-unbounded}
(f_1,\ldots,f_7)
=
(N,N,-N,-N,N,N,N)\,.
\end{equation}
For the $H_3$ ansatz in \eqref{eq:H3-ansatz}, the O1/D1 tadpole condition is then satisfied identically\footnote{It is important to emphasize that the opposite choice $f_i\rightarrow -f_i$ satisfies the tadpole condition equally well. The two choices are related by a skew-whiffing of the $F_5$ flux and differ only in their supersymmetry properties. The scalar potential is insensitive to the overall sign of $F_5$, since it depends quadratically on the fluxes, whereas the Killing-spinor equations are linear in the fluxes and therefore distinguish between the two sign choices. As we will see, only one choice gives rise to the supersymmetric branch.},
\begin{equation}
\sum_{i=1}^{7}h_i f_i
=
-h\left(N+N-N-N\right)
=
0\,,
\end{equation}
and consequently, the flux $N$ is not constrained by tadpole cancellation and may in principle be taken parametrically large.

Having specified a flux configuration that satisfies all tadpole conditions, we
substitute it into the BPS equations in \eqref{eq:BPS-general}. Solving the resulting system, we find that the string coupling is
\begin{align}
e^{\phi}
&=
4\left(
\frac{hN}{\sqrt{3}mq}
\right)^{1/2}
\sim
N^{1/2}q^{-1/2},
\label{eq:dilaton-solution}
\end{align}
where in the last relation we have displayed only the parametric dependence on the fluxes that can be taken unbounded.
The Einstein-frame three-cycle moduli and the $S^1$ radius are given by
\begin{align}
s^1=s^2
&=
\frac{1+\sqrt{3}}{2^{5/4}}
\left(
\frac{\sqrt{3}N q}{mh}
\right)^{3/8}\,,\quad\quad
s^3=s^4
=
\frac{\sqrt{3}-1}{2^{5/4}}
\left(
\frac{\sqrt{3}Nq}{mh}
\right)^{3/8},
\\[2mm]
s^5=s^6=s^7
&=
2^{3/4}
\left(
\frac{\sqrt{3}Nq}{mh}
\right)^{3/8}\,,\quad\quad
r_8
=
\frac{3^{5/16}}{2^{1/4}}
\frac{N^{9/8}}
{m^{1/8}h^{5/8}q^{3/8}}\,.
\label{eq:r8-solution}
\end{align}
Next, to determine the large-volume regime, it is convenient to express the internal radii in string frame.
The Einstein- and string-frame metrics are
related by
\begin{equation}
g^{\rm s}_{MN}=e^{\phi/2}g^E_{MN}\,,
\end{equation}
so that for constant dilaton, the Einstein- and string-frame quantities are related by
\begin{equation}\label{eq:string-Einstein-frame-relations}
r_{\rm s,a}=e^{\phi/4}r_a \,,
\qquad
s^i_{\rm s}=e^{3\phi/4}s^i \,,
\qquad
\mathcal V_{\rm s,7}=e^{7\phi/4}\mathcal V_7\,.
\end{equation}
We first use the relations between the $G_2$ three-cycle moduli and the torus radii, which follow from \eqref{eq:orthonormal-frame}, \eqref{eq:G2-moduli}, and \eqref{eq:integral-basis}, to express the radii in terms of the flux quanta, and then convert them to the string frame. 
We obtain
\begin{equation}\label{eq:string-frame-radii}
\begin{split}
&r_{{\rm s},1}=r_{{\rm s},3}
=
\left(
\frac{2N}{m}
\right)^{1/4}
\sim
N^{1/4}\,,
\qquad
r_{{\rm s},2}=r_{{\rm s},4}
=
\left[
\frac{16(2+\sqrt{3})N}{m}
\right]^{1/4}
\sim
N^{1/4}\,,
\\[2mm]
&r_{{\rm s},5}
=
\left[
\frac{16(2-\sqrt{3})N}{m}
\right]^{1/4}
\sim
N^{1/4}\,,
\qquad
r_{{\rm s},6}
=
(\sqrt{3}-1)
\left(
\frac{N}{2m}
\right)^{1/4}
\sim
N^{1/4}\,,
\\[2mm]
&r_{{\rm s},7}
=
\left[
\frac{(2+\sqrt{3})N}{4m}
\right]^{1/4}
\sim
N^{1/4}\,,
\qquad
r_{{\rm s},8}
=
\left(
\frac{6N^5}{h^2m q^2}
\right)^{1/4}
\sim
N^{5/4}q^{-1/2}\,.
\end{split}
\end{equation}
Thus, $N$ controls the seven-dimensional $G_2$ geometry, while $q$ enters only the string coupling and the radius of the additional $S^1$.

To calculate scale separation, we compare the AdS$_2$ curvature scale with the Kaluza--Klein scales associated with the internal radii. 
Since in two dimensions the vacuum value of the scalar potential does not directly determine the curvature, we first extract $R_2$ from the scalar equations of motion,
\begin{equation}\label{R2}
R_2
=
\left.
\frac{\partial_I V}{\partial_I\mathcal V_8}
\right|_{\rm vac}\,,
\end{equation}
where $I$ runs over all scalar moduli (except the dilaton). 
At the vacuum, each scalar equation must give the same value of $R_2$. 
It is particularly convenient to choose $I=r_8$, since $\mathcal V_8=\mathcal V_7 r_8$ and therefore $\partial_{r_8}\mathcal V_8=\mathcal V_7$. 
We then find
\begin{equation}
R_2
=
\left.
\frac{1}{\mathcal V_7}
\frac{\partial V}{\partial r_8}
\right|_{\rm vac}
=
-\frac{1}{2}\frac{e^{2\phi}m^2}{r_8^2}\,.
\end{equation}

Since the vacuum curvature is negative, we define the Einstein-frame AdS$_2$ radius
\begin{equation}\label{RL}
R_2=-\frac{2}{L_{{\rm AdS}}^{\,2}}\,.
\end{equation}
Since the internal radii have been expressed in string frame, we also convert the AdS radius according to $L_{{\rm AdS},{\rm s}}
=
e^{\phi/4}L_{{\rm AdS}}$, although the comparison of the two scales in Einstein frame leads to the same result.
Substituting the vacuum solution gives
\begin{equation}
L_{{\rm AdS},{\rm s}}
=
\frac{\sqrt{3}}{2^{3/4}}
\frac{N^{3/4}}{h\,m^{3/4}}
\sim
N^{3/4}\,.
\end{equation}
Scale separation requires the Kaluza--Klein scales associated with all internal
directions to be parametrically larger than the AdS$_2$ scale. Equivalently,
we require
\begin{equation}
\frac{L_{{\rm KK},a}^{\,2}}{L_{{\rm AdS},{\rm s}}^{\,2}}
\sim
\frac{r_{{\rm s},a}^{\,2}}{L_{{\rm AdS},{\rm s}}^{\,2}}
\ll 1\,,
\qquad
a=1,\ldots,8\,.
\end{equation}
Using the parametric dependence of the string-frame radii and of the AdS
radius, we find
\begin{align}\label{scaleseparationIIB}
\frac{r_{{\rm s},a}^{\,2}}{L_{{\rm AdS},{\rm s}}^{\,2}}
\sim
\frac{1}{N}\,,
\quad
a=1,\ldots,7,
\quad\quad\quad
\frac{r_{{\rm s},8}^{\,2}}{L_{{\rm AdS},{\rm s}}^{\,2}}
\sim
\frac{N}{q}\,.
\end{align}
To identify classical scale-separated solutions, we require the string-frame radii to be parametrically large and the string coupling to be parametrically small, as in \eqref{eq:string-frame-radii}, while the scale-separation ratios in \eqref{scaleseparationIIB} must be parametrically small, we find
\begin{equation}\label{eq:controlled-regime2}
N\gg1\,,
\qquad
N\ll q\ll N^{5/2}\,.
\end{equation}
The solution admits a parametrically controlled regime in which the string coupling becomes weak, all eight internal string-frame radii become large, and the Kaluza--Klein scales are parametrically separated from the AdS$_2$ scale.
Notice that the upper bound on $q$ follows from requiring the additional $S^1$ radius to remain parametrically large, while the lower bound follows from weak coupling and scale separation along this direction.


\subsection{The O4/O8 system}\label{ssec:O4O8}

The real superpotential reproducing the scalar potential of the massless type IIA
compactification can be written as
\begin{align}\label{WO4general}
W
=
\int_{8}
\Bigg[
&
\left(
\frac12\,\Phi\wedge d\Phi
+
e^{\phi/2}H_{7}
\right)\wedge e^{8}
+
e^{\phi/4}\Psi\wedge F_{4}
+
e^{-3\phi/4}F_{8}
\Bigg] \,.
\end{align}
Substituting \eqref{WO4general} into
\eqref{eq:2d-superpotential-potential} reproduces the scalar potential
obtained from the dimensional reduction,
\begin{equation}
\begin{split}
V
=
&-\int_8 R_7\,\mathrm{vol}_8
+
\frac12\int_8
\left[
e^\phi H_7\wedge\star_8 H_7
+
e^{\phi/2}F_4\wedge\star_8 F_4
+
e^{-3\phi/2}F_8\wedge\star_8 F_8
\right]
\\
&
-
e^{\phi/4}\int_8\Phi\wedge dF_4 \,.
\end{split}
\end{equation}
The last term reproduces the contributions of the smeared O4-plane sources
upon using the Bianchi identity for $F_4$ in
\eqref{F4BianchiO4internal}.

Before turning to specific flux choices, we can integrate the superpotential in
\eqref{WO4general} explicitly over the internal space and obtain the general
superpotential in terms of the fluxes and geometric moduli,
\begin{equation}\label{WO4integrated}
W
=
\frac{r_8}{2}\sum_{i,j=1}^{7}{\cal M}_{ij}s^i s^j
+
e^{\phi/2}q r_8
+
e^{\phi/4}\mathcal V_7
\sum_{i=1}^{7}\frac{f_i}{s^i}
+
e^{-3\phi/4}M \,.
\end{equation}
The supersymmetry conditions are again given by
\eqref{W0}--\eqref{W00}. 
The condition $W_{\phi}=0$ relates the total $F_4$ contribution to those of $H_7$ and $F_8$, while $W=0$ and the seven conditions $W_i=\frac13 W_u$ determine the relations between the geometric contribution and the individual $F_4$ components.
Combining these equations gives
\begin{equation}\label{WO4BPSgeneral}
\begin{split}
e^{\phi/4}\mathcal V_7\frac{f_i}{s^i}
&=
r_8 s^i\sum_{j=1}^{7}{\cal M}_{ij}s^j
+
e^{\phi/2}q r_8\,,
\qquad \text{no sum over }i\,,\\
\frac12\sum_{i,j=1}^{7}{\cal M}_{ij}s^i s^j
&=
-3e^{\phi/2}q \,,\\
M
&=
e^{5\phi/4}q r_8 \,.
\end{split}
\end{equation}


\subsubsection{Example II: Classical solution with scale separation}

We now present a second class of solutions, this time in massless type IIA string theory.
We keep the same Joyce orbifold realization $\mathcal C_{4b}$ on the seven-dimensional space introduced above.
Combining its orbifold elements with the orientifold involution that reflects the additional circle generates four families of spacetime-filling O4-planes, denoted by O4$_{\alpha}$, O4$_{\beta}$, O4$_{\alpha\beta}$, and O4$_{\gamma}$.
For our purposes, we cancel the charge and tension of the O4$_{\alpha}$ and O4$_{\gamma}$ by introducing an appropriate number of D4-branes wrapping the corresponding cycles, leaving O4$_{\beta}$ and O4$_{\alpha\beta}$ as the only non-vanishing net O4 source\footnote{As is clear from the tables in Appendix~\ref{OPlaneconfigurations}, the Joyce orbifold generates O-planes wrapping at least three distinct internal cycles.
We note that configurations with O-planes wrapping only two distinct internal cycles can instead be constructed by modifying the half-period shifts of the orbifold action relative to the standard Joyce realization, as in \cite{VanHemelryck:2025qok}.
In that case, the additional orientifold images can be made freely acting, so that no D-branes are required to cancel the extra O-planes present in the construction used here. With suitable shifts, the O8-plane sector may also be removed entirely.
For convenience, we nevertheless keep the standard Joyce orbifold, since the vacuum properties relevant for our analysis are unchanged at the level of the untwisted truncation.
}.

The two remaining net O4-plane families $O4_{\beta}$ and $O4_{\alpha\beta}$ have geometric involutions on the covering space given by
\begin{equation}
\begin{split}
\sigma_{\mathrm{O4}_{\beta}}
=\sigma_{\mathrm{O8}}\Theta_{\beta}:\quad
(y^{1},\ldots,y^{7},\theta)
&\longmapsto
(-y^{1},-y^{2},y^{3},y^{4},
-y^{5},-y^{6},y^{7},-\theta)\,,
\\[1mm]
\sigma_{\mathrm{O4}_{\alpha\beta}}
=\sigma_{\mathrm{O8}}\Theta_{\alpha}\Theta_{\beta}:\quad
(y^{1},\ldots,y^{7},\theta)
&\longmapsto
(y^{1},y^{2},-y^{3},-y^{4},
-y^{5},-y^{6},y^{7},-\theta)\,.
\end{split}
\end{equation}
At the level of the untwisted truncation, the smeared transverse currents associated with the two remaining net O4-plane families can be chosen as
\begin{equation}
j_{5,\beta}=-\Psi^2\wedge d\theta,
\qquad
j_{5,\alpha\beta}=-\Psi^1\wedge d\theta.
\end{equation}

We now implement the twisting of the torus by replacing the coordinate one-forms $dy^a$ with left-invariant one-forms $\eta^a$ satisfying the Maurer--Cartan equations
\begin{equation}\label{Maurer}
d\eta^a
=
\frac12 f^a{}_{bc}\,
\eta^b\wedge\eta^c\,,
\end{equation}
where $f^a{}_{bc}$ are the quantized structure constants characterizing the metric flux.
We restrict to a single
non-vanishing metric-flux component,
\begin{equation}
d\eta^7=f\,\eta^{56},
\qquad
d\eta^a=0,
\qquad
a=1,\ldots,6\,,
\end{equation}
corresponding to $f^7{}_{56}=f$.
The resulting seven-dimensional space carries a non-trivial $G_2$ structure rather than the torsion-free one considered in the previous example. The metric flux is compatible with the full Joyce orbifold action, since $d\eta^7$ and $\eta^{56}$ transform identically under each of the generators $\Theta_\alpha$, $\Theta_\beta$, and $\Theta_\gamma$. We therefore retain the same notation $\Phi_i$ and $\Psi^i$ for the corresponding orbifold-invariant three- and four-forms.

For the metric flux above, the only non-closed invariant three-forms are
\begin{equation}\label{dPhi1dPhi2}
d\Phi_1=-f\Psi^2,
\qquad
d\Phi_2=-f\Psi^1,
\end{equation}
while the remaining $\Phi_i$ are closed and all $\Psi^i$ remain closed. Consequently, the $G_2$ structure is co-calibrated and satisfies
\begin{equation}
d\Phi=\tau_0\Psi+\star_7\tau_3\,,
\qquad
d\Psi=0\,.
\end{equation}
The explicit torsion classes $\tau_0$ and $\tau_3$ are determined by the exterior derivatives of the $G_2$ structure forms for the chosen metric flux.

We take the NSNS sector to contain an electric three-form flux and its magnetic dual,
\begin{equation}
H_3^{\rm el}
=
E\vol_2\wedge\,d\theta\,,
\qquad
H_7
=
q\,\eta^{1234567}\,,
\end{equation}
which are related by the duality condition \eqref{H7H3}. 
Here $q$ denotes the corresponding quantized flux, while its relation to the electric flux parameter $E$ is given in \eqref{qE}.
The RR fluxes are chosen as
\begin{equation}\label{F44}
F_4
=
\left(
-h\Phi_1-h\Phi_2
+
N\sum_{i=3}^7
\Phi_i \right)\wedge d\theta\,,\quad\quad
F_8
=
M\eta^{1234567}\wedge d\theta \,,
\end{equation}
where $h$, $N$, $q$, and $M$ denote quantized flux parameters. 

The O4-plane tadpoles follow directly from the Bianchi identity for $F_4$. Using the flux ansatz in \eqref{F44} together with the exterior derivatives in \eqref{dPhi1dPhi2}, we find
\begin{align}
dF_4
&=
-h\,d\Phi_1\wedge d\theta
-h\,d\Phi_2\wedge d\theta
\nonumber\\
&=
-hf\left(j_{5,\beta}+j_{5,\alpha\beta}\right) \,.
\end{align}
Thus, for equal net O4 charges $\mu_4<0$, the tadpole condition takes the form
\begin{equation}
-hf=2\kappa_{10}^2\mu_4 \,,
\end{equation}
and we therefore take $f>0, h>0$ so that the two negative O4 charges are canceled by the metric and RR fluxes.
Importantly, the fluxes $N,q,M$ do not enter the tadpole conditions and may therefore be taken parametrically large without requiring cancellations among unbounded fluxes.

The supersymmetry conditions in \eqref{WO4BPSgeneral} imply
\begin{equation}
\begin{aligned}
fs^1s^2
&=
3e^{\phi/2}q\,,
\qquad
\mathcal V_7\frac{h}{s^i}
=
2e^{\phi/4}qr_8,
\quad
i=1,2,
\\[1mm]
M
&=
e^{5\phi/4}qr_8\,,
\qquad
\mathcal V_7\frac{N}{s^i}
=
e^{\phi/4}q r_8,
\quad
i=3,\ldots,7\,.
\end{aligned}
\end{equation}
Solving the equations gives the string coupling
\begin{equation}\label{stringcouplingIIA}
e^\phi
=
\sqrt{\frac{fh}{6}}\,
\frac{M^{3/4}}{q^{1/2}N^{5/4}}\,,
\end{equation}
while the Einstein-frame three-cycle moduli are
\begin{equation}
\begin{split}
s^1
&=
s^2
=
\frac{\sqrt3}{6^{1/8}}
\frac{
h^{1/8}q^{3/8}M^{3/16}
}{
f^{3/8}N^{5/16}
}\,,\quad\quad
r_8
=
6^{5/8}
\frac{
N^{25/16}M^{1/16}
}{
(fh)^{5/8}q^{3/8}
}\,,\\[1mm]
s^i
&=
\frac{2N}{h}s^1\,,\quad\quad i=3\,,\dots ,7 \,.
\end{split}
\end{equation}
Thus all eight geometric moduli as well as the ten-dimensional dilaton are stabilized. 
We use the relations between the $G_2$ three-cycle moduli and the torus radii in \eqref{eq:orthonormal-frame}, \eqref{eq:G2-moduli}, and \eqref{eq:integral-basis}, together with the string- and Einstein-frame relations in \eqref{eq:string-Einstein-frame-relations}, to obtain
\begin{equation}
\begin{split}
r_{{\rm s},1}
&=
r_{{\rm s},2}
=
r_{{\rm s},3}
=
r_{{\rm s},4}
=
\left(\frac{M}{N}\right)^{1/4}\,,\quad\quad
r_{{\rm s},5}
=
r_{{\rm s},6}
=
\sqrt{\frac{2}{h}}\,
(MN)^{1/4}\,,\\[1mm]
r_{{\rm s},7}
&=
\sqrt{\frac{h}{2}}\,
\left(\frac{M}{N^3}\right)^{1/4}\,,\quad\quad
r_{{\rm s},8}
=
\sqrt{\frac{6}{fh}}\,
\frac{M^{1/4}N^{5/4}}{q^{1/2}}
\quad\quad
\mathcal V_{{\rm s},8}
=
\frac{2\sqrt3}{h\sqrt f}\,
\frac{M^2}{q^{1/2}}\,.
\end{split}
\end{equation}
We next determine the AdS$_2$ curvature \eqref{R2} and radius \eqref{RL} and find
\begin{equation}\label{RLIIA}
R_{2,E}
=
-\frac{
f^{9/4}h^{13/4}
}{
6^{9/4}
N^{25/8}q^{1/4}M^{1/8}
}\,,
\qquad
L_{{\rm AdS},{\rm s}}
=
6\sqrt{2}\,
\frac{
N^{5/4}M^{1/4}
}{
f h^{3/2}
}\,,
\end{equation}
and the corresponding geometric scale-separation ratios exhibit the parametric behavior
\begin{equation}\label{scaleseparationIIA}
\frac{r_{{\rm s},1,2,3,4}^2}{L_{{\rm AdS},{\rm s}}^2}
\sim
\frac{1}{N^3}\,,
\qquad
\frac{r_{{\rm s},5,6}^2}{L_{{\rm AdS},{\rm s}}^2}
\sim
\frac{1}{N^2}\,,
\qquad
\frac{r_{{\rm s},7}^2}{L_{{\rm AdS},{\rm s}}^2}
\sim
\frac{1}{N^4}\,,
\qquad
\frac{r_{{\rm s},8}^2}{L_{{\rm AdS},{\rm s}}^2}
\sim
\frac{1}{q}\,.
\end{equation}
Thus, the seven-dimensional geometry becomes parametrically scale separated for large $N$, while scale separation of the additional circle requires large $q$.

We can now determine a simultaneous weak-coupling, large-volume, and scale-separated regime. 
For fixed $f$ and $h$, which are bounded by the O4 tadpole, we find
\begin{equation}\label{eq:controlled-regime22}
N\gg1\,,
\qquad
N^3\ll M\ll N^5\,,
\qquad
M^{3/2}N^{-5/2}
\ll q\ll
M^{1/2}N^{5/2}\,.
\end{equation}
The fluxes $N$, $q$, and $M$, associated respectively with the unbounded component of $F_4$, $H_7$, and $F_8$, do not enter the tadpole conditions and can therefore be taken parametrically large independently. In this regime the string coupling becomes parametrically weak, all eight internal string-frame radii become parametrically large, and all Kaluza--Klein scales are parametrically separated from the AdS$_2$ curvature scale.


\section{Ten-dimensional supersymmetry}
\label{sec:ten-dimensional-supersymmetry}

We now verify directly that the supersymmetry conditions used in Subsection~\ref{ssec:O1O5} follow from the ten-dimensional type IIB Killing-spinor equations. 
We work in the same smeared approximation as in the dimensional reduction, where the ten-dimensional metric is an unwarped direct product, the dilaton and internal metric moduli are constant, and the localized O1/O5 sources are replaced by their uniform distributions.
An analogous calculation can be performed for the type IIA compactification studied in subsection~\ref{ssec:O4O8} but for the ease of presentation we restrict to the type IIB case with an internal $G_2$ holonomy space.

\subsection{Spinors and orientifold projection}

We use the democratic type IIB supersymmetry variations in string frame and
mostly-plus signature, following the conventions reviewed
in~\cite{Koerber:2010bx}. For a differential form $A_p$, we define
$\slashed{A}_p=\frac{1}{p!}A_{M_1\ldots M_p}\Gamma^{M_1\ldots M_p}$,
$H_M=\frac12H_{MNP}\Gamma^{NP}$, and
$\lambda(A_p)=(-1)^{p(p-1)/2}A_p$. The gravitino and dilatino variations are
\begin{align}
\delta\psi^1_M
&=
\left(\nabla_M+\frac14H_M\right)\epsilon_1
+\frac{e^\phi}{16}\slashed{F}_{\rm tot}\Gamma_M\Gamma_{(10)}\epsilon_2,
\nonumber\\
\delta\psi^2_M
&=
\left(\nabla_M-\frac14H_M\right)\epsilon_2
-\frac{e^\phi}{16}\lambda\!\left(\slashed{F}_{\rm tot}\right)
\Gamma_M\Gamma_{(10)}\epsilon_1,
\nonumber\\
\delta\lambda^1
&=
\left(\slashed{\partial}\phi+\frac12\slashed{H}\right)\epsilon_1
+\frac{e^\phi}{16}\Gamma^M\slashed{F}_{\rm tot}
\Gamma_M\Gamma_{(10)}\epsilon_2,
\nonumber\\
\delta\lambda^2
&=
\left(\slashed{\partial}\phi-\frac12\slashed{H}\right)\epsilon_2
-\frac{e^\phi}{16}\Gamma^M
\lambda\!\left(\slashed{F}_{\rm tot}\right)
\Gamma_M\Gamma_{(10)}\epsilon_1,
\label{eq:IIB-supersymmetry-variations}
\end{align}
where $F_{\rm tot}=F_1+F_5+F_9$ in the present truncation and
$\Gamma_{(10)}\epsilon_{1,2}=\epsilon_{1,2}$. Since $\lambda(F_p)=F_p$ for
$p=1,5,9$, the operator $\lambda$ does not introduce an additional sign for
the RR fields retained here.

For the $2+8$ split, we take
\begin{equation}
\Gamma_\mu=\check\gamma_\mu\otimes\mathbf 1,
\qquad
\Gamma_m=\check\gamma_*\otimes\gamma_m,
\qquad
\Gamma_{(10)}=\check\gamma_*\otimes\gamma_9,
\label{eq:10d-gamma-decomposition}
\end{equation}
where $\check\gamma_*^2=\gamma_9^2=1$. A torsion-free $G_2$ metric admits one
covariantly constant real spinor, and its product with the additional circle
gives two covariantly constant eight-dimensional spinors of opposite
chirality,
\begin{equation}
\nabla_m\eta_\pm=0,
\qquad
\gamma_9\eta_\pm=\pm\eta_\pm,
\qquad
\eta_-=\gamma_8\eta_+.
\label{eq:parallel-internal-spinors}
\end{equation}
Before the orientifold projection, the two type IIB Majorana--Weyl parameters
therefore give four real supercharges, corresponding to $\mathcal N=(2,2)$ in
two dimensions.

The orientations of the source currents and the relative RR signs were fixed
in Sections~\ref{sec:orientifold}--\ref{sec:solution}. With these conventions,
the lift of the O1 orientifold action to the ten-dimensional supersymmetry
parameters is
\begin{equation}\label{eq:O1-spinor-projection}
\epsilon_2=-\Gamma_{01}\epsilon_1\,.
\end{equation}
Writing $\check\gamma_*\zeta_\pm=\pm\zeta_\pm$, the compatible decomposition is
\begin{equation}
\begin{aligned}
\epsilon_1
&=
\zeta_+\otimes\eta_+
+
\zeta_-\otimes\eta_-\,,
\\
\epsilon_2
&=
-\zeta_+\otimes\eta_+
+
\zeta_-\otimes\eta_-\,.
\end{aligned}
\label{eq:ten-dimensional-spinor-decomposition}
\end{equation}

Each O5-plane in the configurations of Section~\ref{sec:orientifold} wraps a coassociative four-cycle, which we denote by $\Sigma_{4,i}$.
The source convention in \eqref{eq:smeared-currents} fixes the oriented string-frame volume form of the wrapped cycle to be
\begin{equation}
\vol_{\Sigma_{4,i},{\rm s}}=-\widehat\Psi_i.
\label{eq:O5-cycle-orientation}
\end{equation}
Using the Clifford identities below, the corresponding oriented worldvolume gamma matrix satisfies
\begin{equation}
\Gamma_{\Sigma_{4,i}}\eta_\pm=+\eta_\pm\,.
\label{eq:O5-cycle-Clifford-action}
\end{equation}
The O5 projector is therefore
\begin{equation}
\epsilon_2=-\Gamma_{01}\Gamma_{\Sigma_{4,i}}\epsilon_1\,,
\label{eq:O5-spinor-projection}
\end{equation}
which reduces to the O1 condition \eqref{eq:O1-spinor-projection}.
Thus the O1-plane and all mutually BPS O5-plane families preserve the same two real supercharges, giving $\mathcal N=(1,1)$ supersymmetry; see for example \cite{Koerber:2007jb}.

\subsection{Reduction of the Killing-spinor equations}

The supersymmetry variations are most conveniently evaluated in a string-frame orthonormal basis. 
We introduce the orthonormal forms
\begin{equation}\label{eq:orthonormal-G2-forms}
\widehat\Phi_i=s^i_{\rm s}\Phi_i\,,
\qquad
\widehat\Psi_i=\frac{\mathcal V_{\rm s,7}}{s^i_{\rm s}}\Psi^i\,,
\qquad
\widehat\Psi_i=\star_7\widehat\Phi_i,
\qquad
 e^8_{\rm s}=r_{\rm s,8}d\theta\,,
\end{equation}
For the signs of the basis in \eqref{eq:integral-basis}, and denoting Clifford action by $A\cdot\eta\equiv\slashed A\eta$, the parallel spinors may be chosen such that
\begin{equation}
\gamma_8\eta_\pm=\eta_\mp \,,
\qquad
\widehat\Phi_i\cdot\eta_\pm=\pm\eta_\mp \,,
\qquad
\widehat\Psi_i\cdot\eta_\pm=-\eta_\pm \,.
\label{eq:G2-Clifford-identities}
\end{equation}
In particular, the last identity together with
\eqref{eq:O5-cycle-orientation} gives
\eqref{eq:O5-cycle-Clifford-action}.

Using equations \eqref{F1}-\eqref{F5}, the complete fluxes entering \eqref{eq:IIB-supersymmetry-variations} take the string-frame orthonormal form
\begin{equation}\label{eq:string-frame-fluxes-for-KSE}
\begin{aligned}
F_1
&=
\frac{m}{r_{\rm s,8}}e^8_{\rm s}\,,
\\
H_3
&=
\sum_{i=1}^7\frac{h_i}{s^i_{\rm s}}\widehat\Phi_i
+
q\frac{e^{\phi/4}}{\mathcal V_7}
\vol_{2,\rm s}\wedge e^8_{\rm s}\,,
\\
F_5
&=
\sum_{i=1}^7
\frac{f_i s^i_{\rm s}}{\mathcal V_{\rm s,7}r_{\rm s,8}}
\left(
\widehat\Psi_i\wedge e^8_{\rm s}
-
\vol_{\rm s,2}\wedge\widehat\Phi_i
\right)\,,
\\
F_9
&=
-
\frac{m}{r_{\rm s,8}}\vol_{\rm s,2}\wedge\vol_{\rm s,7}\,.
\end{aligned}
\end{equation}
The relative sign between the two components of $F_5$ follows from
self-duality with the orientation
$\vol_{\rm s,10}=\vol_{\rm s,2}\wedge\vol_{\rm s,7}\wedge e^8_{\rm s}$,
while the last line is the dual of $F_1$ in the democratic formalism. The coefficient of the
electric NSNS component has been written directly in terms of the conserved
$H_7$ flux quantum $q$.

We take the two-dimensional Killing-spinor equation in string frame to be
$\nabla_\mu\zeta_\pm=\kappa_{\rm s}\check\gamma_\mu\zeta_\mp$.
Substituting \eqref{eq:ten-dimensional-spinor-decomposition} and
\eqref{eq:string-frame-fluxes-for-KSE} into the two dilatino variations, the
circle gravitino variation, and the seven $Y_7$ gravitino variations gives,
respectively,
\begin{equation}\label{eq:string-frame-KSE-system}
\begin{split}
\sum_{i=1}^7\frac{h_i}{s^i_{\rm s}}
-
\frac{e^{\phi/4}q}{\mathcal V_7}
&=
-2e^\phi\frac{m}{r_{\rm s,8}}\,,\\
e^\phi
\left(
\sum_{i=1}^7\frac{f_i s^i_{\rm s}}
{\mathcal V_{\rm s,7}r_{\rm s,8}}
-
\frac{m}{r_{\rm s,8}}
\right)
&=
2\frac{e^{\phi/4}q}{\mathcal V_7}\,,\\
-\frac{e^\phi}{6}
\left(
\frac{m}{r_{\rm s,8}}
+
\sum_{j=1}^7\frac{f_j s^j_{\rm s}}
{\mathcal V_{\rm s,7}r_{\rm s,8}}
-
6\frac{f_i s^i_{\rm s}}
{\mathcal V_{\rm s,7}r_{\rm s,8}}
\right)
&=
\frac{h_i}{s^i_{\rm s}} \,,
\qquad i=1,\ldots,7.
\end{split}
\end{equation}
Summing the last line and combining the result with the first two equations
yields
\begin{equation}
\sum_{i=1}^7
\frac{f_i s^i_{\rm s}}{\mathcal V_{\rm s,7}r_{\rm s,8}}
=
2\frac{m}{r_{\rm s,8}}\,,
\qquad
\frac{e^{\phi/4}q}{\mathcal V_7}
=
\frac{e^\phi}{2}\frac{m}{r_{\rm s,8}}\,,
\qquad
\frac{h_i}{s^i_{\rm s}}
=
e^\phi
\left(
\frac{f_i s^i_{\rm s}}{\mathcal V_{\rm s,7}r_{\rm s,8}}
-
\frac{m}{2r_{\rm s,8}}
\right)\,.
\label{eq:string-frame-KSE-solution}
\end{equation}
Converting with \eqref{eq:string-Einstein-frame-relations}, the independent
ten-dimensional supersymmetry conditions become
\begin{equation}
\begin{aligned}
e^{-\phi/2}\mathcal V_8\frac{h_i}{s^i}
-
f_i s^i
&=
-\frac12e^\phi m\mathcal V_7,
\qquad \text{no sum over }i\,,\\
\sum_{i=1}^7f_i s^i
&=
2e^\phi m\mathcal V_7,
\\
e^{\phi/2}q r_8
&=
\frac12e^\phi m\mathcal V_7.
\end{aligned}
\label{eq:ten-dimensional-BPS-conditions}
\end{equation}
These equations agree exactly with \eqref{eq:BPS-general}, obtained from the
real superpotential in Section~\ref{sec:solution}. Thus the supersymmetry
conditions used there are the complete bulk ten-dimensional Killing-spinor
equations for the constant-dilaton, unwarped ansatz.

The external gravitino variation gives the Einstein-frame Killing-spinor
equation
\begin{equation}
\nabla_\mu\zeta_\pm
=
\frac{e^\phi m}{4r_8}
\check\gamma_\mu\zeta_\mp\,.
\label{eq:AdS2-Killing-spinor-equation}
\end{equation}
Its integrability condition fixes the curvature and the AdS$_2$ radius,
\begin{equation}\label{eq:Lambda-Einstein-BPS}
R_{2,E}
=
-\frac12e^{2\phi}\frac{m^2}{r_8^2}\,,
\qquad
L_{\rm AdS}
=
\frac{2r_8}{e^\phi|m|}\,.
\end{equation}
This agrees with the curvature obtained independently from the scalar
equations in Section~\ref{sec:solution}. Since the quantized $H_7$ flux $q$
is the conserved electric displacement of the two-dimensional field
$B_{\mu\theta}$, equation~\eqref{eq:ten-dimensional-BPS-conditions} gives
\begin{equation}\label{eq:q-BPS}
q
=
\frac12e^{\phi/2}m\frac{\mathcal V_7}{r_8}\,.
\end{equation}
In particular, $m>0$ requires $q>0$ on the supersymmetric branch.

\subsection{The explicit \texorpdfstring{$\mathcal C_{4b}$}{C4b} solution}

We now apply the ten-dimensional equations to the explicit vacuum of Section~\ref{sec:solution}, with the flux choices \eqref{eq:H3-ansatz} and \eqref{eq:F5-unbounded}. We take $m,h,N>0$, and hence $q>0$. The flux ansatz and the BPS equations are invariant under the permutations $1\leftrightarrow2$, $3\leftrightarrow4$, and $5\leftrightarrow6\leftrightarrow7$, thus the $i=5,6,7$ equations in \eqref{eq:ten-dimensional-BPS-conditions} immediately give $s^5=s^6=s^7$. 
The $i=3,4$ equations are the same quadratic equation with a unique positive root, so $s^3=s^4$.
The $i=1,2$ equations are likewise the same quadratic equation.
If $s^1$ and $s^2$ were its two distinct positive
roots, their sum would contradict the second equation in \eqref{eq:ten-dimensional-BPS-conditions} for $s^3>0$, hence $s^1=s^2$.
Denoting the common value of $s^5,s^6,s^7$ by $s^5$, the $i=5,6,7$ equations give
\begin{equation}
N s^5
=
\frac12e^\phi m\mathcal V_7 \,.
\label{eq:KSE-i567}
\end{equation}
The sum equation and the $i=1,3$ equations then reduce to
\begin{equation}\label{eq:KSE-shape-relations}
2\left(s^1-s^3\right)=s^5\,,
\qquad
s^1\left(s^5-s^1\right)=s^3\left(s^5+s^3\right)\,,
\end{equation}
whose positive solution is
\begin{equation}\label{eq:KSE-shape-solution}
s^1=s^2=\frac{1+\sqrt3}{4}s^5\,,
\qquad
s^3=s^4=\frac{\sqrt3-1}{4}s^5\,,
\qquad
s^5=s^6=s^7\,.
\end{equation}
and it follows that $\mathcal V_7=\frac14(s^5)^{7/3}$. The remaining shape equation, the sum equation, and the electric equation become
\begin{equation}
e^{-\phi/2}\mathcal V_7r_8h
=
\frac{\sqrt3}{8}N(s^5)^2,
\qquad
e^\phi m\mathcal V_7
=
2Ns^5,
\qquad
e^{\phi/2}q r_8
=
Ns^5.
\label{eq:KSE-scale-equations}
\end{equation}
Solving these equations gives
\begin{equation}\label{eq:KSE-explicit-solution}
\begin{split}
e^\phi
&=
4\left(
\frac{hN}{\sqrt3mq}
\right)^{1/2} \,,
\\
s^5=s^6=s^7
&=
2^{3/4}
\left(
\frac{\sqrt3Nq}{mh}
\right)^{3/8} \,,
\\
r_8
&=
\frac{3^{5/16}}{2^{1/4}}
\frac{N^{9/8}}{m^{1/8}h^{5/8}q^{3/8}}\,.
\end{split}
\end{equation}
Together with \eqref{eq:KSE-shape-solution}, this reproduces precisely the solution in \eqref{eq:dilaton-solution}--\eqref{eq:r8-solution}. Therefore, the scale-separated family constructed in Section~\ref{sec:solution} contains a ten-dimensional $\mathcal N=(1,1)$ supersymmetric branch.

\section{\texorpdfstring{Geometric scale separation and obstruction}{}}
\label{sec:ads2-no-go}

The supersymmetric solutions constructed above provide a direct test of the obstruction to scale-separated $\mathcal N\geq(1,1)$ $\mathrm{AdS}_2$ flux vacua proposed in \cite{Cribiori:2024jwq}.
It is important to distinguish two notions of scale separation.
In the geometric sense used in this work, all internal Kaluza--Klein scales are parametrically above the $\mathrm{AdS}_2$ scale, as shown in \eqref{scaleseparationIIB} for type IIB and \eqref{scaleseparationIIA} for type IIA.
Restricting to the controlled regimes in \eqref{eq:controlled-regime2} and \eqref{eq:controlled-regime22}, respectively, we find
\begin{equation}
m_{\rm KK}L_{\rm AdS}\rightarrow\infty\,,
\end{equation}
so that the Kaluza--Klein modes decouple from physics at the $\mathrm{AdS}_2$ curvature scale in the corresponding large-flux limits.
The argument of \cite{Cribiori:2024jwq}, by contrast, concerns the existence of a Wilsonian two-dimensional theory with a cutoff parametrically above $L_{\rm AdS}^{-1}$ and allows the scale obstructing such a description to be non-geometric.
For the type IIB solution, combining \eqref{eq:Lambda-Einstein-BPS} and \eqref{eq:q-BPS} gives the relation below, while for the type IIA solution the same relation follows directly from the explicit vacuum solution
\begin{equation}\label{eq:ads2-f1-bps-relation}
\frac{1}{L_{\rm AdS}}
=
q\frac{e^{\phi/2}}{\mathcal V_7}\,.
\end{equation}

The natural BPS domain wall in both compactifications is a fundamental string\footnote{Both theories contain other domain walls such as Dp-branes wrapping internal $p$-cycles.} wrapped on the internal $S^1$. It appears as a particle in two dimensions, is electrically charged under $B_{\mu\theta}$, and changes $q$, the conserved electric displacement and equivalently the quantized $H_7$ flux, by one unit. 
To make this relation explicit, consider a fundamental string wrapped on the
$S^1$ and localized along the spatial direction of $\mathrm{AdS}_2$. 
Its Einstein-frame action is
\begin{equation}
S_{\rm F1}
=
-T_{\rm F1}\int d^{10}X\,\sqrt{-g_{10}}\,
e^{\phi/2}\rho_{\rm F1}
+
T_{\rm F1}\int B_2\wedge j_8\,,
\end{equation}
where $T_{\rm F1}=(2\pi\alpha')^{-1}$, while after smearing the string over $Y_7$ and keeping it localized at $x^1=x_0$ in the spatial direction of $\mathrm{AdS}_2$, its source density is
\begin{equation}
\rho_{\rm F1}
=
\frac{1}{\mathcal V_7}
\frac{\delta(x^1-x_0)}{\sqrt{g_{11}}}\,.
\end{equation}
Since the wrapped F1 string extends along $t$ and $\theta$, variation of its DBI action gives the non-vanishing mixed components $T^0{}_0=T^\theta{}_\theta
=
-T_{\rm F1}e^{\phi/2}\rho_{\rm F1}$, with all transverse components vanishing. 
Contracting the trace-reversed ten-dimensional Einstein equation over the two external directions, the contribution of the wrapped fundamental string to the two-dimensional curvature is therefore
\begin{equation}
\left.R_2\right|_{\rm F1}
=
-\frac12
\left(
\kappa_{10}^2 T_{\rm F1}
\frac{e^{\phi/2}}{\mathcal V_7}
\right)
\frac{\delta(x^1-x_0)}{\sqrt{g_{11}}}\,,\qquad
T_{\rm F1}^{(2)}
\equiv
\kappa_{10}^2 T_{\rm F1}
\frac{e^{\phi/2}}{\mathcal V_7}\,,
\end{equation}
with $T_{\rm F1}^{(2)}$ the gravitationally normalized two-dimensional tension.
Combining \eqref{eq:ads2-f1-bps-relation} with the definition above gives
\begin{equation}\label{TF12}
T_{\rm F1}^{(2)}
\sim
\frac{1}{qL_{\rm AdS}}\,.
\end{equation}
If the tension of this fundamental BPS wall is assumed to provide an upper bound on the Wilsonian cutoff, then
\begin{equation}\label{eq:cribiori-cutoff}
\Lambda_{\rm UV}
\lesssim
T_{\rm F1}^{(2)}
\sim
\frac{1}{qL_{\rm AdS}}
\ll
\frac{1}{L_{\rm AdS}}
\ll
m_{\rm KK}\,,
\end{equation}
so that the actual cutoff may be lower, but cannot be parametrically higher than the BPS domain-wall scale.
The solutions therefore do not define parametrically gapped autonomous two-dimensional Wilsonian theories in the sense of~\cite{Cribiori:2024jwq}. 
Thus, although the Kaluza--Klein tower decouples parametrically, the wrapped string introduces an additional non-geometric BPS scale that prevents the two-dimensional theory from having a cutoff parametrically above the $\mathrm{AdS}_2$ scale.

This conclusion does not mean that the wrapped string has a small local rest mass. From the smeared ten-dimensional F1 action above, the rest mass of a string wrapping once around $S^1_\theta$ is $m_{\rm F1}^{\rm rest}=T_{\rm F1}e^{\phi/2}r_8$.
Using the large-flux scalings we find
\begin{equation}
m_{\rm F1}^{\rm rest}L_{\rm AdS}
\sim
N^2q^{-1/2}
\longrightarrow\infty\,,
\end{equation}
which is guaranteed throughout the controlled regime \eqref{eq:controlled-regime2}.
The wrapped string is therefore parametrically heavy in ordinary rest-mass units.
The quantity in \eqref{TF12}, by contrast, measures the strength with which the wrapped string sources the two-dimensional curvature after smearing over the internal space.
Thus, a microscopically heavy object can nevertheless introduce a parametrically low gravitationally normalized BPS scale without becoming a light neutral perturbative particle.

There is also a fixed-charge subtlety. The two-dimensional potential was obtained by integrating out the electric field at fixed quantized displacement $q$, whereas a single wrapped fundamental string acts on the Hilbert space schematically as $\mathcal H_q \longrightarrow \mathcal H_{q\pm1}$, and strings with higher wrapping numbers connect sectors whose charges differ by more than one unit. Such objects therefore do not furnish a tower of neutral low-dimension operators in the operator algebra of a single fixed-charge CFT$_1$ rather interpolate between two distinct AdS$_2$ solutions with correspondingly distinct dual CFT$_1$'s.

The same distinction is visible after T-duality along the circle in the smeared background. The electric field of $B_{\mu\theta}$ becomes the electric field of a KK graviphoton.
Working in string frame, the T-dual radius and the corresponding dimensionless electric field, which takes the critical value for the present BPS background, are
\begin{equation}
r'_8=\frac{\alpha'}{r_{{\rm s},8}},
\qquad
E_{\rm KK}=\frac{L_{{\rm AdS},{\rm s}}}{r'_8}\,.
\end{equation}
More generally, consider a three-dimensional scalar of mass $M_{(3)}$ and expand it in modes with momentum $n$ along the fiber. The resulting two-dimensional field has electric charge $n$ and obeys
\begin{equation}
\begin{split}
\Delta_n(\Delta_n-1)
&=
\left(
M_{(3)}^2+\frac{n^2}{(r'_8)^2}
\right)L_{{\rm AdS},{\rm s}}^2
-
n^2E_{\rm KK}^2
\\
&=
M_{(3)}^2L_{{\rm AdS},{\rm s}}^2.
\end{split}
\label{eq:ads2-charged-dimension-cancellation}
\end{equation}
For the pure momentum state T-dual to the ground state fundamental string wrapped on $S^1_\theta$ one has $M_{(3)}=0$. 
The KK momentum contribution is then canceled completely by the coupling to the critical electric field, giving formally $\Delta_n(\Delta_n-1)=0$.

We therefore conclude that the solutions are genuinely geometrically scale separated and admit a classical two-dimensional dilaton-supergravity reduction in the smeared approximation. Via the wrapped fundamental strings, they also realize the BPS flux-changing objects predicted in~\cite{Cribiori:2024jwq}. Under the additional assumption that the fundamental string's gravitationally normalized tension bounds the cutoff of the full two-dimensional theory, the solutions are compatible with the proposed obstruction and do not yield an autonomous parametrically gapped Wilsonian $\mathrm{AdS}_2$ theory. However, whether this conclusion must also apply to the neutral sector at fixed electric charge is not clear to us, and we leave this question for future work.


\section{T-dualities}\label{sec:Dualities}

We discuss the T-dualities of the type IIB compactification with O1/O5-planes and fluxes studied above for the orientifold system $\mathcal C_{4b}$, and also include the corresponding dualities for the same compactification on $\mathcal C_{4a}$. 
Although the underlying $\mathcal C_{4a}$ and $\mathcal  C_{4b}$ Joyce orbifolds are related by an affine redefinition of the torus coordinates and a relabeling of the orbifold generators, this transformation does not map the unshifted O1 involution used here to itself.
The resulting orientifold and flux backgrounds are therefore inequivalent and give rise to distinct networks of geometric T-dualities.

Denoting by $T_a$ a T-duality along the direction $y^a$, the condition for a T-duality to remain geometric follows from the standard chain \cite{Shelton:2005cf}
\begin{equation}
    H_{abc}
    \xrightarrow{\,T_a\,}
    f^a{}_{bc}
    \xrightarrow{\,T_b\,}
    Q^{ab}{}_{c}
    \xrightarrow{\,T_c\,}
    R^{abc}\,,
    \label{HfQRchain}
\end{equation}
where $Q^{ab}{}_{c}$ and $R^{abc}$ correspond to non-geometric fluxes\footnote{Performing a second T-duality along another leg of the original $H_{abc}$ component produces $Q$-flux, while dualizing also the third leg produces $R$-flux. 
These cases are non-geometric in the sense that they cannot be described globally in terms of an ordinary metric and $B$ field on a conventional manifold.}.
The metric flux arises when a T-duality is performed along one of the directions threaded by a component of $H_3$ and describes the twisting of the corresponding circle according to \eqref{Maurer}.
A component $H_{abc}$ therefore remains geometric provided that at most one of the directions $a,b,c$ is T-dualized. 
If none of them is dualized, the component remains as ordinary $H_3$ flux, while if exactly one is dualized it is converted into metric flux.

Considering all of the above, we perform multiple T-dualities under the requirement that the dual theory remains \textit{geometric}.
\begin{table}[ht]
\centering
\small
\renewcommand{\arraystretch}{1.35}
\setlength{\tabcolsep}{5pt}
\begin{tabular}{|c|c|c|c|c|}
\hline
&
T-dualities
&
structure of $X_8$
&
RR sector
&
O-plane content
\\
\hline

\multirow{7}{*}{$\mathcal C_{4b}$}
&
$T_1,\ T_3,\ T_6$
&
$SU(3)\times S^1\times S^1_\theta$
&
$F_2+F_4+F_6$
&
O2+2\,O4+2\,O6
\\[1mm]

&
$T_2,\ T_4,\ T_5$
&
$SU(3)\times S^1\times S^1_\theta$
&
$F_2+F_4+F_6$
&
O2+3\,O4+O6
\\[1mm]

&
$T_7$
&
$SU(3)\times S^1\times S^1_\theta$
&
$F_2+F_4+F_6$
&
O2+O4+3\,O6
\\[1mm]

\cline{2-5}

&
$T_{14},\ T_{15},\ T_{23},\ T_{26},\ T_{35},\ T_{46}$
&
$SU(3)\times S^1\times S^1_\theta$
&
$F_3+F_5+F_7$
&
2\,O3+3\,O5
\\[1mm]

&
$T_{24},\ T_{25},\ T_{45}$
&
$SU(3)\times S^1\times S^1_\theta$
&
$F_3+F_5+F_7$
&
3\,O3+2\,O5
\\[1mm]

\cline{2-5}

&
$T_{145},\ T_{235},\ T_{246}$
&
$G_2\times S^1_\theta$
&
$F_4+F_8$
&
5\,O4
\\[1mm]

&
$T_{245}$
&
$SU(3)\times S^1\times S^1_\theta$
&
$F_2+F_4+F_6$
&
O2+4\,O4
\\[1mm]

\hline

\multirow{5}{*}{$\mathcal C_{4a}$}
&
$T_1,\ T_2,\ T_3,\ T_5,\ T_6,\ T_7$
&
$SU(3)\times S^1\times S^1_\theta$
&
$F_2+F_4+F_6$
&
O2+2\,O4+2\,O6
\\[1mm]

&
$T_4$
&
$SU(3)\times S^1\times S^1_\theta$
&
$F_2+F_4+F_6$
&
O2+4\,O4
\\[1mm]

\cline{2-5}

&
$T_{15},\ T_{26},\ T_{37}$
&
$SU(3)\times S^1\times S^1_\theta$
&
$F_3+F_5+F_7$
&
O3+4\,O5
\\[1mm]

&
$T_{14},\ T_{24},\ T_{34},\ T_{45},\ T_{46},\ T_{47}$
&
$SU(3)\times S^1\times S^1_\theta$
&
$F_3+F_5+F_7$
&
3\,O3+2\,O5
\\[1mm]

\cline{2-5}

&
$T_{145},\ T_{246},\ T_{347}$
&
$G_2\times S^1_\theta$
&
$F_4+F_8$
&
5\,O4
\\[1mm]

\hline
\end{tabular}
\caption{Details of the internal spaces obtained from the geometric T-dualities of the $\mathcal C_{4a}$ and $\mathcal C_{4b}$ configurations of the type IIB O1/O5 setup. Here $S^1_{\theta}$ denotes the circle of the original geometry, which is not involved in the T-dualities. The coefficients in the last column count orientifold families rather than individual fixed loci.}
\label{Tdualitystructuretable}
\end{table}
Starting from our setup with $G_2$ holonomy, we find that T-dualities reduce the structure group of the dual geometries according to
\begin{equation}
G_2
\supset
SU(3)\,.
\end{equation}
A seven-dimensional $SU(3)$ structure is specified by a globally defined one-form $v$, together with a two-form $J$ and a complex three-form $\Omega$ defined on the six-dimensional distribution orthogonal to $v$.
The $G_2$ structure forms can then be decomposed in terms of these quantities as \cite{Chiossi:2002tw}
\begin{equation}\label{G2splitSingleT}
    \Phi
    =
    v\wedge J+\operatorname{Re}\Omega\,,
    \qquad
    \Psi
    =
    \frac12 J\wedge J
    -v\wedge\operatorname{Im}\Omega\,.
\end{equation}

After a single T-duality along $y^a$, the dualized direction defines the one-form $v$ of the resulting $SU(3)$ structure. 
For two T-dualities along $y^a$ and $y^b$, the one-form $v$ is instead associated with the third direction $y^c$ that completes the corresponding associative triple, namely $\iota_b\iota_a\Phi=\pm v$.
When all three directions of an associative triple are T-dualized, the resulting geometry again carries a $G_2$ structure, whereas a triple T-duality along directions that do not form an associative triple leads instead to an $SU(3)$ structure.
The resulting configurations are presented in Table~\ref{Tdualitystructuretable}, while further details on the dual geometries, including the NSNS sector, are provided in Appendix~\ref{sec:Tdualitiesdetails}.

Substituting the $SU(3)$ structure decomposition \eqref{G2splitSingleT} into the general superpotential \eqref{WIIgeneral}, and using $J\wedge\Omega=0$, the geometric contribution becomes
\begin{equation}
\frac12\,\Phi\wedge d\Phi
=
\frac12\,v\wedge dv\wedge J^2
+
v\wedge\operatorname{Re}\Omega\wedge dJ
+
\frac12\,\operatorname{Re}\Omega\wedge d\operatorname{Re}\Omega\,.
\end{equation}
When $d\operatorname{Re}\Omega$ has no component along $v$, the last term vanishes. If $v$ is moreover an untwisted spectator direction, so that $dv=0$, the first term vanishes as well. For the single T-dualities, however, $v$ generally parameterizes the circle twisted by the metric flux, so that $dv\neq0$ and the first term must be retained.
The corresponding even and odd RR calibration polyforms, obtained by T-duality of the
$G_2$ expressions above, are
\begin{equation}
\begin{split}
\widetilde{\mathcal C}_{\rm even}^{SU(3)}
&=
\widetilde{\mathcal C}_2+\widetilde{\mathcal C}_4+\widetilde{\mathcal C}_6
\\
&=
\left(v\wedge e^8+J\right)
+
\left(
v\wedge\operatorname{Re}\Omega
-e^8\wedge\operatorname{Im}\Omega
\right)
-
\left(
\frac12\,v\wedge e^8\wedge J^2
+\frac16 J^3
\right),
\\[2mm]
\widetilde{\mathcal C}_{\rm odd}^{SU(3)}
&=
\widetilde{\mathcal C}_1+\widetilde{\mathcal C}_3+\widetilde{\mathcal C}_5+\widetilde{\mathcal C}_7
\\
&=
v
+
\left(
-\operatorname{Im}\Omega
+e^8\wedge J
\right)
+
\left(
\frac12\,v\wedge J^2
-v\wedge e^8\wedge\operatorname{Re}\Omega
\right)
+\frac16\,e^8\wedge J^3\,.
\end{split}
\end{equation}

For the T-dual configurations, and especially for the single T-dualities, it is useful to allow for a more general flux ansatz rather than restricting to an isotropic one.
As explained earlier, additional independent fluxes lead to radii that depend on a larger set of flux parameters, thereby allowing one to explore broader families of solutions.

As we have seen in T-duals of AdS$_3$ compactifications on $G_2$ orbifolds \cite{Miao:2025rgf,Tringas:2026ncg}, the dual geometries can involve co-calibrated $G_2$ structures as well as $SU(3)$ structure spaces times a spectator circle. An important feature of these constructions is the anisotropy of the internal geometry induced by the independent flux parameters. Keeping a sufficiently general flux ansatz allows the different internal radii to depend on different combinations of flux quanta and therefore gives access to a broad range of parametric regimes, including large or small individual radii, weak or strong coupling, and scale separation.
Since the T-dual theories of our setup are closely related to the AdS$_3$ compactifications, but contain additional fluxes subject to further constraints, such as the external flux, it would be interesting to investigate which families of solutions can be realized and whether they exhibit the same range of parametric regimes.


\section{Conclusion}\label{sec:conclusion}

To the best of our knowledge, flux compactifications of type II string theory to two dimensions on seven-dimensional $G_2$ structure spaces times a circle with \textit{smeared} orientifold sources have not previously been studied.
In this work, we perform a general analysis of two classes of such compactifications, a type IIB setup with O1/O5-planes and a massless type IIA setup with O4/O8-planes.
In both cases, the orientifold projections leave a sufficiently rich flux sector to stabilize the untwisted geometric moduli and the ten-dimensional dilaton, while allowing flux directions that are not bounded by the tadpole conditions and lead to parametrically controlled classical $\mathrm{AdS}_2$ vacua with geometric scale separation.

More specifically, motivated by the different mechanisms of tadpole cancellation, which can leave distinct imprints on the conformal spectrum of the putative holographic dual, we studied two different classes of solutions.
For the type IIB O1/O5 system, we found a family of $\mathrm{AdS}_2$ vacua in which an unbounded $F_5$ flux together with the magnetic NSNS flux $H_7$ controls the parametric regime. Taking the corresponding flux quanta large within the appropriate window leads simultaneously to weak string coupling, parametrically large internal radii in string units, and a parametrically large hierarchy between all Kaluza--Klein scales and the $\mathrm{AdS}_2$ curvature scale. We derived the corresponding two-dimensional dilaton-gravity theory and its real superpotential, and verified directly from the ten-dimensional type IIB Killing-spinor equations that the scale-separated family contains a branch preserving $\mathcal N=(1,1)$ supersymmetry. The ten-dimensional conditions reproduce precisely those obtained from the two-dimensional superpotential. Based on this setup, we also studied its geometric T-dualities and classified the resulting dual theories with geometric backgrounds, which are compactifications on either $G_2$ structure spaces times a circle or $SU(3)$ structure spaces times a two-torus.

We also constructed a second class of solutions in massless type IIA with O4/O8-planes. In this case, the O4-plane tadpoles are supported by metric flux together with non-closed components of $F_4$, while the remaining $F_4$ fluxes, $H_7$, and $F_8$ are unconstrained by the tadpoles. These unbounded flux directions again allow a simultaneous weak-coupling, large-volume, and scale-separated regime. This provides a qualitatively different mechanism from the type IIB construction and shows that parametrically controlled $\mathrm{AdS}_2$ vacua are not restricted to a single orientifold or flux configuration.

These supersymmetric two-dimensional solutions also provide a useful setting in which to compare geometric scale separation with the proposed obstruction to scale-separated $\mathrm{AdS}_2$ vacua with extended supersymmetry in \cite{Cribiori:2024jwq}. 
Although all geometric Kaluza--Klein scales are parametrically above the $\mathrm{AdS}_2$ scale, the wrapped fundamental strings that change the quantized $H_7$ flux have a gravitationally normalized tension with precisely the scaling expected from the proposed obstruction. 
Our solutions are therefore compatible with this obstruction despite being geometrically scale separated. 
At the same time, since these objects interpolate between sectors with different electric charge, it remains an open question whether the same conclusion applies to the neutral spectrum within a fixed-charge sector.

A natural first step toward understanding the holographic interpretation of these vacua would be to compute the conformal dimensions of the dual operators and test the recent holographic constraints \cite{Bobev:2025yxp,Revello:2026eqp} on scale-separated AdS vacua, with the aim of determining whether and in what way the resulting two-dimensional theories differ from their higher-dimensional counterparts with similar parametric properties.

\subsection*{Acknowledgments}
We would like to thank Niccolò Cribiori and Thomas Van Riet for discussions. 
This work is supported in part by the NSF grant PHY-2609419 and the Lehigh University CORE grant with grant ID COREAWD40.

\paragraph*{AI assistance.}
The authors used ChatGPT-5.6 Sol Pro and Claude Opus 5.0 Max for language editing and to check calculations. The authors take full responsibility for the content and accuracy of the work.

\appendix


\section{O-plane configurations}\label{OPlaneconfigurations}

In this Appendix, we discuss the different orientifold configurations arising from different choices of shifts in the Joyce orbifold defined in \eqref{orbifold}--\eqref{orbifold2}.

Among the configurations in Table~\ref{tab:Oplane-shifts}, $\mathcal C_{3a}$ and $\mathcal C_{3b}$ correspond to Joyce orbifolds for which smooth $G_2$ resolutions are known \cite{Joyce:1996a,Joyce:1996b}.
The configuration $\mathcal C_{3b}$ is the standard Joyce example, for which the singular locus consists of twelve disjoint three-tori with local geometry $T^3\times\mathbb C^2/\mathbb Z_2$. 
These singularities can be resolved by replacing the $\mathbb C^2/\mathbb Z_2$ factor by an Eguchi--Hanson space, leading to a smooth manifold of $G_2$ holonomy. 
The configuration $\mathcal C_{3a}$ is a closely related Joyce example obtained by removing one of the half-period shifts.
In this case the singular locus contains eight $T^3$ components together with eight $T^3/\mathbb Z_2$ components that are locally tensored with $\mathbb C^2/\mathbb Z_2$.
The corresponding resolution is more involved but is likewise part of the Joyce construction. 
\begin{table}[ht]
\centering
\small
\renewcommand{\arraystretch}{1.15}
\setlength{\tabcolsep}{5pt}

\begin{tabular}{|c|c|ccccccc|}

\multicolumn{9}{c}{\textit{O1/O5 system}}
\\
\hline
Configuration
&
$(c_1,c_2,c_3)$
&
O5$_{\alpha}$
&
O5$_{\alpha\gamma}$
&
O5$_{\alpha\beta}$
&
O5$_{\beta}$
&
O5$_{\alpha\beta\gamma}$
&
O5$_{\beta\gamma}$
&
O5$_{\gamma}$
\\
\hline
$\mathcal C_{3a}$ &
$(\frac12,\frac12,0)$ &
$\checkmark$ & $\checkmark$ & $\checkmark$ & -- & -- & -- & --
\\
$\mathcal C_{3b}$ &
$(\frac12,\frac12,\frac12)$ &
$\checkmark$ & $\checkmark$ & $\checkmark$ & -- & -- & -- & --
\\
$\mathcal C_{4a}$ &
$(\frac12,0,\frac12)$ &
$\checkmark$ & $\checkmark$ & $\checkmark$ & -- & $\checkmark$ & -- & --
\\
$\mathcal C_{4b}$ &
$(0,\frac12,\frac12)$ &
$\checkmark$ & $\checkmark$ & $\checkmark$ & $\checkmark$ & -- & -- & --
\\
$\mathcal C_{5a}$ &
$(\frac12,0,0)$ &
$\checkmark$ & $\checkmark$ & $\checkmark$ & -- & $\checkmark$ & -- & $\checkmark$
\\
$\mathcal C_{5b}$ &
$(0,\frac12,0)$ &
$\checkmark$ & $\checkmark$ & $\checkmark$ & $\checkmark$ & -- & $\checkmark$ & --
\\
$\mathcal C_{5c}$ &
$(0,0,\frac12)$ &
$\checkmark$ & $\checkmark$ & $\checkmark$ & $\checkmark$ & $\checkmark$ & -- & --
\\
$\mathcal C_{7}$ &
$(0,0,0)$ &
$\checkmark$ & $\checkmark$ & $\checkmark$ & $\checkmark$ &
$\checkmark$ & $\checkmark$ & $\checkmark$
\\
\hline
\multicolumn{9}{c}{}
\\[-2mm]

\multicolumn{9}{c}{\textit{O4/O8 system}}
\\
\hline
Configuration
&
$(c_1,c_2,c_3)$
&
O4$_{\alpha}$
&
O4$_{\alpha\gamma}$
&
O4$_{\alpha\beta}$
&
O4$_{\beta}$
&
O4$_{\alpha\beta\gamma}$
&
O4$_{\beta\gamma}$
&
O4$_{\gamma}$
\\
\hline
$\mathcal C_{3a}$ &
$(\frac12,\frac12,0)$ &
$\checkmark$ & -- & -- & $\checkmark$ & -- & -- & $\checkmark$
\\
$\mathcal C_{3b}$ &
$(\frac12,\frac12,\frac12)$ &
$\checkmark$ & -- & -- & $\checkmark$ & -- & -- & $\checkmark$
\\
$\mathcal C_{4a}$ &
$(\frac12,0,\frac12)$ &
$\checkmark$ & -- & -- & $\checkmark$ & -- & $\checkmark$ & $\checkmark$
\\
$\mathcal C_{4b}$ &
$(0,\frac12,\frac12)$ &
$\checkmark$ & -- & $\checkmark$ & $\checkmark$ & -- & -- & $\checkmark$
\\
$\mathcal C_{5a}$ &
$(\frac12,0,0)$ &
$\checkmark$ & $\checkmark$ & -- & $\checkmark$ & -- & $\checkmark$ & $\checkmark$
\\
$\mathcal C_{5b}$ &
$(0,\frac12,0)$ &
$\checkmark$ & -- & $\checkmark$ & $\checkmark$ & $\checkmark$ & -- & $\checkmark$
\\
$\mathcal C_{5c}$ &
$(0,0,\frac12)$ &
$\checkmark$ & -- & $\checkmark$ & $\checkmark$ & -- & $\checkmark$ & $\checkmark$
\\
$\mathcal C_{7}$ &
$(0,0,0)$ &
$\checkmark$ & $\checkmark$ & $\checkmark$ & $\checkmark$ &
$\checkmark$ & $\checkmark$ & $\checkmark$
\\
\hline
\end{tabular}

\caption{Orientifold configurations for the O1/O5 and O4/O8 systems for different choices of half-period shifts $(c_1,c_2,c_3)$. 
A check mark denotes an orientifold-plane family with a fixed locus.}
\label{tab:Oplane-shifts}
\end{table}

The $\mathcal C_{4a}$, $\mathcal C_{4b}$ and $\mathcal C_{5a}$, $\mathcal C_{5b}$, $\mathcal C_{5c}$ configurations have a more involved singularity structure. 
For these choices of shifts, additional non-trivial elements of the orbifold group possess fixed loci and some of the corresponding singular sets intersect.
At the resulting loci, the local geometry can be of the form $\mathbb R\times\mathbb C^3/(\mathbb Z_2\times\mathbb Z_2)$, rather than the standard singularity of the simplest Joyce construction. 
Such singularities can nevertheless be treated within the more general Joyce construction. 
In particular, the transverse $\mathbb C^3/(\mathbb Z_2\times\mathbb Z_2)$ quotient admits crepant resolutions carrying Ricci-flat QALE metrics with holonomy $SU(3)$ \cite{joyce2000compact,joyce1999quasialemetricsholonomysum}, which provide the local building blocks for resolving the compact $G_2$ orbifold.

The unshifted configuration $\mathcal C_7$ is the most singular case within the family considered here.
All seven non-trivial elements of $\Gamma$ possess fixed loci, leading to a large number of intersecting singular sets.
The standard Joyce resolution does not directly apply to this configuration, and the resolution of the resulting orbifold singularities is not well understood.

Representative configurations from each of the three classes in Table~\ref{tab:Oplane-shifts} discussed above have already appeared in three-dimensional compactifications.
Compactifications based on the Joyce resolved configuration $\mathcal C_{3a}$ were studied in \cite{Farakos:2025bwf,Miao:2025rgf}. 
In those constructions, stabilization of all moduli required a non-vanishing net contribution from D-branes to the scalar potential, however, this is not the case for the other two classes.
The configuration $\mathcal C_{4a}$ was considered in \cite{Tringas:2026ncg}, where the D-brane and orientifold contributions can cancel, so that no net D-brane contribution to the scalar potential is required for moduli stabilization. 
The maximally singular configuration $\mathcal C_7$ was studied in \cite{Farakos:2020phe}, where moduli stabilization can likewise be achieved without a net D-brane contribution. 
This suggests that the richer fixed-locus and flux structure of the latter configurations provides additional freedom in satisfying the tadpole conditions and stabilizing the moduli without a net D-brane contribution, with the source sector contributing only through a net orientifold term.
Nevertheless, in all three cases supersymmetric vacua with stabilized untwisted moduli and parametric scale separation were found, and at the level of the untwisted sector they share broadly similar qualitative properties.


\section{Two-dimensional vacuum equations}\label{2Dequation}

We derive the conditions for maximally symmetric vacua with constant scalar fields. 
Denoting collectively the scalar moduli by $\varphi^I$, the terms relevant for such backgrounds take the form
\begin{equation}\label{eq:2d-vacuum-action}
S_2
=
\frac{1}{2\kappa_{10}^{2}}
\int d^2x\,\sqrt{-g_2}
\left(
\mathcal V_8 R_2 - V
\right)\,,
\end{equation}
where both $\mathcal V_8$ and $V$ depend on the scalar fields $\varphi^I$. 
The scalar kinetic terms have been omitted here since they vanish for constant-field configurations.

Because the coefficient of the Ricci scalar is field dependent, variation
with respect to a scalar modulus gives
\begin{equation}\label{eq:general-scalar-eom}
\partial_I V
=
R_2\,\partial_I\mathcal V_8\,.
\end{equation}
The metric equation following from \eqref{eq:2d-vacuum-action} is
\footnote{
Here $\nabla_\mu$ denotes the covariant derivative associated with the two-dimensional metric $g_{\mu\nu}$, while $\Box\equiv g^{\rho\sigma}\nabla_\rho\nabla_\sigma$ is the corresponding covariant d'Alembertian.
Since $\mathcal V_8$ is a scalar, $\nabla_\mu\mathcal V_8=\partial_\mu\mathcal V_8$, whereas $\nabla_\mu\nabla_\nu\mathcal V_8= \partial_\mu\partial_\nu\mathcal V_8-\Gamma^\rho_{\mu\nu}\partial_\rho\mathcal V_8$.
The derivative terms in \eqref{eq:2d-metric-eom} arise from the variation of the non-minimal coupling $\mathcal V_8 R_2$.}
\begin{equation}\label{eq:2d-metric-eom}
\mathcal V_8 G_{\mu\nu}
+
\left(
g_{\mu\nu}\Box-\nabla_\mu\nabla_\nu
\right)\mathcal V_8
+
\frac12 V g_{\mu\nu}
=
0\,.
\end{equation}
However, in two dimensions the Ricci tensor satisfies
\begin{equation}
R_{\mu\nu}
=
\frac12 R_2 g_{\mu\nu}\,,
\end{equation}
and therefore the Einstein tensor vanishes identically,
\begin{equation}
G_{\mu\nu}
=
R_{\mu\nu}
-\frac12 R_2 g_{\mu\nu}
\equiv 0 \,.
\end{equation}
Moreover, for a vacuum with constant scalar moduli, $\partial_\mu\varphi^I=0$ we see from \eqref{eq:general-scalar-eom} that $\mathcal V_8$ is constant in spacetime, consequently 
\begin{equation}
\Box\mathcal V_8
=
\nabla_\mu\nabla_\nu\mathcal V_8
=
0\,.
\end{equation}
The metric equation \eqref{eq:2d-metric-eom} therefore reduces to
\begin{equation}
\frac12 V g_{\mu\nu}=0\,,
\end{equation}
which, for a non-degenerate metric, implies the potential $V$ to be zero.

It is important to emphasize that $V=0$ does not imply a vanishing cosmological constant. 
The reduced theory is a two-dimensional dilaton-gravity theory in which the internal volume $\mathcal V_8$ multiplies the Ricci scalar. Unlike in dimensions $D>2$, this field-dependent prefactor cannot be removed by a conventional Weyl rescaling to an ordinary Einstein frame.
Consequently, the value of the scalar potential at the vacuum is not directly identified with the cosmological constant.

For a maximally symmetric two-dimensional spacetime we define
\begin{equation}\label{eq:2d-curvature}
R_{\mu\nu}
=
\Lambda g_{\mu\nu}\,,
\qquad
R_2=2\Lambda\,,
\end{equation}
and therefore the scalar equations \eqref{eq:general-scalar-eom} become
\begin{equation}\label{eq:general-vacuum-eom}
\partial_I V
=
2\Lambda\,\partial_I\mathcal V_8\,.
\end{equation}
Thus the curvature is determined by the scalar equations rather than by the value of $V$ itself.
In particular, an AdS$_2$ vacuum is characterized by $\Lambda<0$, while simultaneously satisfying $V=0$.


\section{Superpotential/potential details}\label{sec:superpotentialdetails}

We provide some more information on the derivation of the scalar potential using the superpotential in \eqref{eq:2d-superpotential-potential}.
For completeness, we give the inverse of the scalar field-space metric introduced in \eqref{KIJ}. In the coordinates \eqref{moduli1}-\eqref{moduli2}, the non-vanishing components of the inverse metric are
\begin{equation}
K^{ij}
=
\frac{4}{7\mathcal V_8}
\left(1-7\delta^{ij}\right)\,,\quad\quad
K^{iu}
=
\frac{6}{7\mathcal V_8}\,,\quad\quad
K^{uu}
=
-\frac{12}{7\mathcal V_8}\,,\quad\quad
K^{\phi\phi}
=
-\frac{4}{\mathcal V_8}\,,
\end{equation}
with $K^{i\phi}=K^{u\phi}=0$. The indices $i,j$ label the $G_2$ moduli, while $u$ denotes the logarithmic modulus associated with the additional $S^1$ direction.
For the gravitational coupling function in \eqref{J}, the corresponding derivatives are
\begin{equation}
J_i=\frac{2}{3}\mathcal V_8\,,
\qquad
J_u=2\mathcal V_8\,,
\qquad
J_\phi=0\,.
\end{equation}
Contracting these expressions with the inverse field-space metric gives
\begin{equation}
J_IK^{Ii}=\frac{12}{7}\,,
\qquad
J_IK^{Iu}=\frac{4}{7}\,,
\qquad
J_IK^{I\phi}=0\,,
\qquad
J_IK^{IJ}J_J
=
\frac{64}{7}\mathcal V_8\,.
\end{equation}

It is useful to express the superpotential in \eqref{eq:superpotential} as
\begin{equation}
W=A+B+C-D\,,
\end{equation}
where
\begin{align}
A
&=
\sum_{i=1}^7 A_i
=
e^{-\phi/2}\int_{8} H_3\wedge\Psi\wedge e^8
=
e^{-\phi/2}
\sum_{i,j}
h_i\,\frac{\mathcal V_7}{s^j}\,r_8
\int \Phi_i\wedge\Psi^j\wedge d\theta
=
e^{-\phi/2}\mathcal V_8
\sum_i\frac{h_i}{s^i}\,,
\nonumber\\
B
&=
e^{-\phi/2}\int_{8} e^\phi H_7\wedge e^8
=
e^{\phi/2}q\,r_8
\int_{8}dy^{1234567}\wedge d\theta
=
e^{\phi/2}qr_8\,,
\nonumber\\
C
&=
\sum_{i=1}^7 C_i
=
\int_{8}\Phi\wedge F_5
=
\sum_{i,j}s^i f_j
\int_{8}\Phi_i\wedge\Psi^j\wedge d\theta
=
\sum_{i=1}^{7}f_i s^i\,,
\nonumber\\
D
&=
e^\phi\int_{8}\mathrm{vol}_7\wedge F_1
=
e^\phi m
\int_{8}\mathrm{vol}_7\wedge d\theta
=
e^\phi m\mathcal V_7\,.
\label{ABCD}
\end{align}
Using these definitions, the quantities entering
\eqref{eq:2d-superpotential-potential} are
\begin{align}
J_IK^{IJ}W_J
&=
\frac{20}{7}A
+\frac{4}{7}B
+\frac{12}{7}C
-4D\,,
\nonumber\\
4W-J_IK^{IJ}W_J
&=
\frac{8}{7}\left(A+3B+2C\right)\,,
\\
K^{IJ}W_IW_J
&=
\frac{1}{7\mathcal V_8}
\Bigg[
A^2
+6AB
+4AC
-56AD
-19B^2
+12BC
+4C^2
-28D^2
\nonumber\\
&\quad\quad\quad
-28\sum_{i=1}^{7}A_i^2
+56\sum_{i=1}^{7}A_iC_i
-28\sum_{i=1}^{7}C_i^2
\Bigg]\,.
\nonumber
\end{align}
Substituting these expressions into
\eqref{eq:2d-superpotential-potential}, the terms proportional to
$A^2$, $C^2$, $AC$, $AB$, and $BC$ cancel, and we obtain
\begin{equation}\label{eq:potentialABCD}
V
=
\frac{1}{\mathcal V_8}
\left[
\frac{1}{2}\sum_{i=1}^{7}A_i^2
+
\frac{1}{2}B^2
+
\frac{1}{2}\sum_{i=1}^{7}C_i^2
+
\frac{1}{2}D^2
+
AD
-
\sum_{i=1}^{7}A_iC_i
\right].
\end{equation}
To rebuild the dimensionally reduced scalar potential in differential form as in
\eqref{eq:potential-forms}, we use the orientation in \eqref{orientation}.
For a \(p\)-form \(\alpha_p\) supported entirely on \(Y_7\), the
seven- and eight-dimensional Hodge stars are related by
\begin{equation}
\star_8\alpha_p
=
(\star_7\alpha_p)\wedge e^8,
\qquad
\star_8(\alpha_p\wedge e^8)
=
(-1)^{7-p}\star_7\alpha_p\,.
\end{equation}
In particular,
\begin{equation}
\star_8e^8=-\vol_7,
\qquad
\star_8\vol_7=e^8,
\qquad
\star_8\Phi_i=(\star_7\Phi_i)\wedge e^8,
\qquad
\star_8(\Psi^i\wedge e^8)=-\star_7\Psi^i\,,
\end{equation}
and we use the normalization
\begin{equation}
\int_{X_8}\Phi_i\wedge\Psi^j\wedge d\theta
=
\delta_i{}^j\,.
\end{equation}
Using \eqref{eq:7d-hodge} together with \eqref{eq:orthonormal-frame},
the four quadratic terms in the potential become
\begin{equation}
\begin{split}
\frac{1}{2\mathcal V_8}A_i^2
&=
\frac12 e^{-\phi}
\mathcal V_7r_8
\frac{h_i^2}{(s^i)^2}
=
\frac12 e^{-\phi}
\int_{8}
H_{3,i}\wedge\star_8 H_{3,i}\,,\\
\frac{1}{2\mathcal V_8}B^2
&=
\frac12 e^\phi q^2\frac{r_8}{\mathcal V_7}
=
\frac12 e^\phi
\int_{8}
H_7\wedge\star_8 H_7 \,, \\
\frac{1}{2\mathcal V_8}C_i^2
&=
\frac12
\frac{f_i^2(s^i)^2}{\mathcal V_7r_8}
=
\frac12
\int_{8}
F_{5,i}^{\rm mag}\wedge\star_8 F_{5,i}^{\rm mag}\,,\\
\frac{1}{2\mathcal V_8}D^2
&=
\frac12 e^{2\phi}m^2\frac{\mathcal V_7}{r_8}
=
\frac12 e^{2\phi}
\int_{8}
F_1\wedge\star_8 F_1\,.
\end{split}
\end{equation}
Here the fluxes are expanded on the eight-dimensional internal space as in
\eqref{H3FULL}, \eqref{F1}, and \eqref{F5}, with no sum over \(i\).

The two remaining terms in \eqref{eq:potentialABCD} are the cross terms between the NSNS and RR contributions.
Using \eqref{ABCD}, they can be written as
\begin{equation}
\begin{split}
\frac{AD}{\mathcal V_8}
&=
e^{\phi/2}m\mathcal V_7
\sum_{i=1}^{7}\frac{h_i}{s^i}
=
e^{\phi/2}
\int_{8}
\Psi\wedge H_3^{\rm int}\wedge F_1
=
2\kappa_{10}^2
\sum_{i=1}^{7}
\mu_{5,i}\,
e^{\phi/2}
\frac{\mathcal V_7}{s^i}\,,
\\
-\frac{1}{\mathcal V_8}
\sum_{i=1}^{7}A_iC_i
&=
-e^{-\phi/2}
\sum_{i=1}^{7}h_if_i
=
-e^{-\phi/2}
\int_{8}
H_3^{\rm int}\wedge F_5^{\rm mag}
=
2\kappa_{10}^2
\mu_1e^{-\phi/2}\,.
\end{split}
\end{equation}
In the last equalities, we used the Bianchi identities
\eqref{eq:F7-Bianchi-direct} together with the definitions of the smeared currents.
These terms reproduce the O5/D5 and O1/D1 contributions to the scalar potential,
respectively. Combining the quadratic and cross terms then reproduces the
differential-form expression for the scalar potential in
\eqref{eq:potential-forms}.


\section{T-duality details}\label{sec:Tdualitiesdetails}

In this Appendix, we present further details of the T-dualities discussed in Section~\ref{sec:Dualities} for which the dual background remains geometric.
\begingroup
\small
\renewcommand{\arraystretch}{1.1}
\setlength{\tabcolsep}{7pt}

\begin{longtable}{|c|c|c|r@{\,$=$\,}l|}
\hline
&
T-duality
&
$H_3$
&
\multicolumn{2}{|c|}{metric fluxes}
\\
\hline
\endfirsthead

\hline
&
T-duality
&
$H_3$
&
\multicolumn{2}{|c|}{metric fluxes}
\\
\hline
\endhead

\hline
\endfoot

\endlastfoot


\multicolumn{5}{c}{\textit{Single T-dualities}}
\\
\hline

\multirow{7}{*}{$\mathcal C_{4b}$}
&
$T_1$
&
$h(\eta^{347}+\eta^{567})$
&
$\left(f^1{}_{27},f^1{}_{36}\right)$
&
$(-h,-h)$
\\*

&
$T_2$
&
$-h(-\eta^{347}-\eta^{567}+\eta^{136})$
&
$f^2{}_{17}$
&
$h$
\\*

&
$T_3$
&
$-h(\eta^{127}-\eta^{567})$
&
$\left(f^3{}_{47},f^3{}_{16}\right)$
&
$(h,h)$
\\*

&
$T_4$
&
$-h(\eta^{127}-\eta^{567}+\eta^{136})$
&
$f^4{}_{37}$
&
$-h$
\\*

&
$T_5$
&
$-h(\eta^{127}-\eta^{347}+\eta^{136})$
&
$f^5{}_{67}$
&
$h$
\\*

&
$T_6$
&
$-h(\eta^{127}-\eta^{347})$
&
$\left(f^6{}_{57},f^6{}_{13}\right)$
&
$(-h,-h)$
\\*

&
$T_7$
&
$-h\,\eta^{136}$
&
$\left(f^7{}_{12},f^7{}_{34},f^7{}_{56}\right)$
&
$(-h,h,h)$
\\

\hline

\multirow{7}{*}{$\mathcal C_{4a}$}
&
$T_1$
&
$h(\eta^{567}+\eta^{235})$
&
$\left(f^1{}_{27},f^1{}_{36}\right)$
&
$(-h,-h)$
\\*

&
$T_2$
&
$-h(-\eta^{567}+\eta^{136})$
&
$\left(f^2{}_{17},f^2{}_{35}\right)$
&
$(h,h)$
\\*

&
$T_3$
&
$-h(\eta^{127}-\eta^{567})$
&
$\left(f^3{}_{25},f^3{}_{16}\right)$
&
$(-h,h)$
\\*

&
$T_4$
&
$-h(\eta^{127}-\eta^{567}+\eta^{136}-\eta^{235})$
&
$f^4{}_{bc}$
&
$0$
\\*

&
$T_5$
&
$-h(\eta^{127}+\eta^{136})$
&
$\left(f^5{}_{67},f^5{}_{23}\right)$
&
$(h,h)$
\\*

&
$T_6$
&
$-h(\eta^{127}-\eta^{235})$
&
$\left(f^6{}_{57},f^6{}_{13}\right)$
&
$(-h,-h)$
\\*

&
$T_7$
&
$-h(\eta^{136}-\eta^{235})$
&
$\left(f^7{}_{12},f^7{}_{56}\right)$
&
$(-h,h)$
\\


\hline
\multicolumn{5}{c}{\textit{Double T-dualities}}
\\
\hline

\multirow{9}{*}{$\mathcal C_{4b}$}
&
$T_{14}$
&
$h\,\eta^{567}$
&
$\left(f^1{}_{27},f^1{}_{36},f^4{}_{37}\right)$
&
$(-h,-h,-h)$
\\*

&
$T_{15}$
&
$h\,\eta^{347}$
&
$\left(f^1{}_{27},f^1{}_{36},f^5{}_{67}\right)$
&
$(-h,-h,h)$
\\*

&
$T_{23}$
&
$h\,\eta^{567}$
&
$\left(f^2{}_{17},f^3{}_{47},f^3{}_{16}\right)$
&
$(h,h,h)$
\\*

&
$T_{24}$
&
$-h(-\eta^{567}+\eta^{136})$
&
$\left(f^2{}_{17},f^4{}_{37}\right)$
&
$(h,-h)$
\\*

&
$T_{25}$
&
$-h(-\eta^{347}+\eta^{136})$
&
$\left(f^2{}_{17},f^5{}_{67}\right)$
&
$(h,h)$
\\*

&
$T_{26}$
&
$h\,\eta^{347}$
&
$\left(f^2{}_{17},f^6{}_{57},f^6{}_{13}\right)$
&
$(h,-h,-h)$
\\*

&
$T_{35}$
&
$-h\,\eta^{127}$
&
$\left(f^3{}_{47},f^3{}_{16},f^5{}_{67}\right)$
&
$(h,h,h)$
\\*

&
$T_{45}$
&
$-h(\eta^{127}+\eta^{136})$
&
$\left(f^4{}_{37},f^5{}_{67}\right)$
&
$(-h,h)$
\\*

&
$T_{46}$
&
$-h\,\eta^{127}$
&
$\left(f^4{}_{37},f^6{}_{57},f^6{}_{13}\right)$
&
$(-h,-h,-h)$
\\

\hline

\multirow{9}{*}{$\mathcal C_{4a}$}
&
$T_{15}$
&
$0$
&
$\left(f^1{}_{27},f^1{}_{36},f^5{}_{67},f^5{}_{23}\right)$
&
$(-h,-h,h,h)$
\\*

&
$T_{26}$
&
$0$
&
$\left(f^2{}_{17},f^2{}_{35},f^6{}_{57},f^6{}_{13}\right)$
&
$(h,h,-h,-h)$
\\*

&
$T_{37}$
&
$0$
&
$\left(f^3{}_{25},f^3{}_{16},f^7{}_{12},f^7{}_{56}\right)$
&
$(-h,h,-h,h)$
\\*

&
$T_{14}$
&
$h(\eta^{567}+\eta^{235})$
&
$\left(f^1{}_{27},f^1{}_{36}\right)$
&
$(-h,-h)$
\\*

&
$T_{24}$
&
$-h(-\eta^{567}+\eta^{136})$
&
$\left(f^2{}_{17},f^2{}_{35}\right)$
&
$(h,h)$
\\*

&
$T_{34}$
&
$-h(\eta^{127}-\eta^{567})$
&
$\left(f^3{}_{25},f^3{}_{16}\right)$
&
$(-h,h)$
\\*

&
$T_{45}$
&
$-h(\eta^{127}+\eta^{136})$
&
$\left(f^5{}_{67},f^5{}_{23}\right)$
&
$(h,h)$
\\*

&
$T_{46}$
&
$-h(\eta^{127}-\eta^{235})$
&
$\left(f^6{}_{57},f^6{}_{13}\right)$
&
$(-h,-h)$
\\*

&
$T_{47}$
&
$-h(\eta^{136}-\eta^{235})$
&
$\left(f^7{}_{12},f^7{}_{56}\right)$
&
$(-h,h)$
\\


\hline
\multicolumn{5}{c}{\textit{Triple T-dualities}}
\\
\hline

\multirow{4}{*}{$\mathcal C_{4b}$}
&
$T_{145}$
&
$0$
&
$\left(f^1{}_{27},f^1{}_{36},f^4{}_{37},f^5{}_{67}\right)$
&
$(-h,-h,-h,h)$
\\*

&
$T_{235}$
&
$0$
&
$\left(f^2{}_{17},f^3{}_{47},f^3{}_{16},f^5{}_{67}\right)$
&
$(h,h,h,h)$
\\*

&
$T_{245}$
&
$-h\,\eta^{136}$
&
$\left(f^2{}_{17},f^4{}_{37},f^5{}_{67}\right)$
&
$(h,-h,h)$
\\*

&
$T_{246}$
&
$0$
&
$\left(f^2{}_{17},f^4{}_{37},f^6{}_{57},f^6{}_{13}\right)$
&
$(h,-h,-h,-h)$
\\

\hline

\multirow{3}{*}{$\mathcal C_{4a}$}
&
$T_{145}$
&
$0$
&
$\left(f^1{}_{27},f^1{}_{36},f^5{}_{67},f^5{}_{23}\right)$
&
$(-h,-h,h,h)$
\\*

&
$T_{246}$
&
$0$
&
$\left(f^2{}_{17},f^2{}_{35},f^6{}_{57},f^6{}_{13}\right)$
&
$(h,h,-h,-h)$
\\*

&
$T_{347}$
&
$0$
&
$\left(f^3{}_{25},f^3{}_{16},f^7{}_{12},f^7{}_{56}\right)$
&
$(-h,h,-h,h)$
\\

\hline

\caption{Geometric single, double, and triple T-dualities of the
$\mathcal C_{4a}$ and $\mathcal C_{4b}$ backgrounds.
The last column lists the non-vanishing metric-flux components
$f^a{}_{bc}$ generated by the corresponding T-dualities.
The electric NSNS component, or equivalently its magnetic dual $H_7$,
is present in all cases.}
\label{Tdualitygeometrytable}
\\

\end{longtable}

\endgroup

\bibliographystyle{JHEP}
\bibliography{refs}

\end{document}